\documentclass[floats,floatfix,showpacs,amssymb,prd,twocolumn,superscriptaddress,nofootinbib]{revtex4-1}
\usepackage{amssymb,amsmath,verbatim,color,mathtools,enumitem,xcolor}
\usepackage[normalem]{ulem}
\usepackage{hyperref}
 \definecolor{linkcolor}{rgb}{0.0,0.3,0.5}
 \hypersetup{colorlinks=true,linkcolor=linkcolor,citecolor=linkcolor,filecolor=linkcolor,urlcolor=linkcolor}
\usepackage{graphicx}

\definecolor{dwycomment}{HTML}{1F77B4}

\newcommand\unit[1]{{\rm #1}}

\newcommand\skipme[1]{}

\newcommand\skipDavideFormat[1]{}

\newcommand\TdaysOone{48.6}    %
\newcommand\TdaysAnalyzedOtwoEvent{11}

\newcommand{\mc}{\mathcal{M}_{\mathrm{c}}}
\newcommand\E[1]{{\left<#1\right>}}

\def\apj{\rm{ApJ}}
\def\apjl{\rm{ApJ}}
\def\apjs{\rm{ApJS}}

\def\aap{\rm{A\&A}}

\def\mnras{\rm{MNRAS}}
\def\na{\rm{New A}}

\def\prd{\rm{PRD}}

\def\prl{\rm{PRL}}

\def\nat{\rm{Nature}}

\def\cqg{\rm{CQG}}
\def\prx{\rm{PRX}}

\newcommand{\approptoinn}[2]{\mathrel{\vcenter{
  \offinterlineskip\halign{\hfil$##$\cr
    #1\propto\cr\noalign{\kern2pt}#1\sim\cr\noalign{\kern-2pt}}}}}

\begin{document}

\title{Explaining LIGO's observations via isolated binary evolution with natal kicks}
\author{Daniel Wysocki}
\email{dw2081@rit.edu}
\affiliation{Rochester Institute of Technology, Rochester, New York 14623, USA}
\author{Davide Gerosa}
\affiliation{TAPIR 350-17, California Institute of Technology, 1200 E California Boulevard, Pasadena, California 91125, USA}
\author{Richard O'Shaughnessy}
\affiliation{Rochester Institute of Technology, Rochester, New York 14623, USA}
\author{Krzysztof Belczynski}
\affiliation{Nicolaus Copernicus Astronomical Centre, Polish Academy of Sciences, Ulica Bartycka 18, 00-716 Warsaw, Poland}
\author{Wojciech Gladysz}
\affiliation{Astronomical Observatory, Warsaw University, Aleje Ujazdowskie 4, 00-478 Warsaw, Poland}
\author{Emanuele Berti}
\affiliation{Department of Physics and Astronomy, The University of Mississippi, University, Mississippi 38677, USA}
\affiliation{CENTRA, Departamento de F\'isica, Instituto Superior T\'ecnico, Universidade de Lisboa, Avenida Rovisco Pais 1, 1049 Lisboa, Portugal}
\author{Michael Kesden}
\affiliation{Department of Physics, The University of Texas at Dallas, Richardson, Texas 75080, USA}
\author{Daniel E. Holz}
\affiliation{Enrico Fermi Institute, Department of Physics, Department of Astronomy and Astrophysics, and Kavli Institute for Cosmological Physics, University of Chicago, Chicago, Illinois 60637, USA}
\affiliation{Kavli Institute for Particle Astrophysics \& Cosmology and Physics Department, Stanford University, Stanford, California 94305, USA}


\date{\today}

\frenchspacing

\begin{abstract}
We compare binary evolution models with different assumptions about black-hole natal kicks to the first gravitational-wave observations performed by the LIGO detectors.  Our comparisons attempt to reconcile merger rate, masses, spins, and spin-orbit misalignments of all current observations with state-of-the-art formation scenarios of binary black holes formed in isolation.  We estimate that black holes (BHs) should receive natal kicks at birth of the order of $\sigma\simeq 200$ (50) km/s if tidal processes do (not) realign stellar spins.  Our estimate is driven by two simple factors.  The natal kick dispersion $\sigma$ is bounded from above because large kicks disrupt too many binaries (reducing the merger rate below the observed value).  Conversely, the natal kick distribution is bounded from below because modest kicks are needed to produce a range of spin-orbit misalignments.  A distribution of misalignments increases our models' compatibility with LIGO's observations, if all BHs are likely to have natal spins.  Unlike related work which adopts a concrete BH natal spin prescription, we explore a range of possible BH natal spin distributions.  Within the context of our models, for all of the choices of $\sigma$ used here and within the context of one simple fiducial parameterized spin distribution, observations favor low BH natal spin.
\end{abstract}

\maketitle

\section{Introduction}

The  discovery and interpretation of gravitational waves (GW) from coalescing binaries \cite{2016PhRvL.116f1102A} has initiated a revolution in astronomy \cite{2016ApJ...818L..22A}.  Several hundred more detections are expected over the next five years  \cite{2016ApJ...833L...1A,2016PhRvX...6d1015A,2017PhRvL.118v1101A}.
Already, the properties of the sources responsible---the inferred event rates, masses, and spins---have confronted other observations of black hole (BH) masses and spins \cite{2016PhRvX...6d1015A}, challenged previous formation scenarios \cite{2016ApJ...818L..22A,2016PhRvX...6d1015A}, and inspired new models \cite{2016MNRAS.458.2634M,2016A&A...588A..50M,2016ApJ...824L...8R,2016PhRvL.116t1301B} and insights \cite{2016MNRAS.462..844K,2016MNRAS.463L..31L} into the evolution of massive stars and the observationally accessible gravitational waves they emit \cite{2016MNRAS.461.3877D,2016PhRvL.116m1102A}.  Over the next several years, our understanding of the lives and deaths of massive stars over cosmic time will be transformed by the identification and interpretation of the population(s) responsible for coalescing binaries, with and without counterparts, because  measurements will enable  robust tests to distinguish between formation scenarios with present \cite{2016ApJ...832L...2R,2017PhRvL.119a1101O} and future instruments \cite{2016ApJ...830L..18B,2016PhRvD..94f4020N}, both coarsely and with high precision.
In this work, we demonstrate the power of gravitational wave measurements to constrain how BHs form, within the context of one formation scenario for binary BHs: the isolated evolution of pairs of stars \cite{1993MNRAS.260..675T,1997MNRAS.288..245L,2002ApJ...572..407B,2001A&A...375..890N,2003MNRAS.342.1169V,2012ApJ...759...52D,2013ApJ...779...72D,2015ApJ...806..263D,2016Natur.534..512B,2016MNRAS.462.3302E,2017NatCo...814906S}.
Within the context of that model, we focus our attention on the one feature whose unique impacts might be most observationally accessible: BH natal kicks.  Observations strongly suggest that when compact objects like neutron stars are formed after the death of a massive star, their birth can impart significant linear momentum or ``kick.''  For example, observations of pulsars in our galaxy suggest birth velocity changes  as high as $v_k\sim 450$ km/s \cite{2005MNRAS.360..974H}.  These impulsive momentum changes impact the binary's intrinsic orbit and stability, changing the orbital parameters like semimajor axis and orbital plane \cite{2013PhRvD..87j4028G,2000ApJ...541..319K}, as well as causing the center of mass of the remnant BH binary (if still bound) to recoil at a smaller but still appreciable velocity.  While no single compelling and unambiguous  observation can be explained only with a BH natal kick, the assumption of small but nonzero BH natal kicks provides a natural explanation for several observations, including the posterior spin-orbit misalignment distribution of GW151226 and the galactic x-ray binary misalignment \cite{1995Natur.375..464H,1997ApJ...477..876O,2001ApJ...555..489O,2010MNRAS.401.1514M} and recoil velocity \cite{2012MNRAS.425.2799R,2015MNRAS.453.3341R,2017MNRAS.467..298R,2016ApJ...819..108B,2012ApJ...747..111W,2014ApJ...790..119W}.  Modest BH natal kicks can be produced by, for example, suitable neutrino-driven supernova engines; see, e.g., \cite{2017arXiv170607053B} and references therein.

We compare binary formation models with different BH natal kick prescriptions to LIGO observations of binary black holes.  Along with \cite{2017arXiv170607053B}, our calculation is one of the first to perform this comparison while changing a single, physically well-defined and astrophysically interesting parameter: the BH natal kick strength.  It is the first to self-consistently draw inferences about binary evolution physics by comparing observations simultaneously to the predicted detection
rate; binary BH masses; and binary BH spins, accounting for both magnitude and misalignment.

This comparison is important because BH natal kicks introduce two complementary and unusually distinctive effects on the binary BHs that LIGO detects.  On the one hand, strong BH natal kicks will frequently disrupt possible progenitor binary systems.  As the strength of BH natal kicks increases, the expected number of coalescing binary BHs drops precipitously \cite{1997MNRAS.288..245L,1999A&A...346...91B,2002ApJ...572..407B}.  On the basis of observations to date, BH natal kicks drawn from a distribution with one-dimensional velocity dispersion $\sigma $ greater than $265~\unit{km/s}$ are disfavored \cite{2016Natur.534..512B}.
On the other hand, BH natal kicks will tilt the orbital plane, misaligning the orbital angular momentum from the black hole's natal spin direction---assumed parallel to the progenitor binary's orbital angular momentum \cite{2000ApJ...541..319K,2017PhRvL.119a1101O}.
The imprint of these natal kicks on the binary's dynamics is preserved over the aeons between the BH-BH binary's formation and its final coalescence \cite{2008PhRvD..78d4021R,2013PhRvD..87j4028G,2015PhRvL.114h1103K,2015PhRvD..92f4016G}.  The outgoing radiation from each merger contains information about the coalescing binary's spin (see, e.g., \cite{2015PhRvD..91d2003V,2016PhRvL.116x1102A,2016PhRvD..94f4035A} and references therein), including conserved constants that directly reflect the progenitor binary's state \cite{2016PhRvD..93d4071T,2014PhRvD..89l4025G}.  Several studies have demonstrated that the imprint of processes that misalign BH spins and the orbit can be disentangled \cite{2017CQGra..34cLT01V,2017MNRAS.471.2801S,2017Natur.548..426F}.

In this work, we show that LIGO's observations of binary black holes can be easily explained in the context of isolated binary evolution, if BH natal kicks act with the (modest) strength to misalign the orbital plane from the initial spin directions (presumed aligned).  In this approach, the absence of large aligned spins either reflects fortuitous but nonrepresentative observations or low natal BH spins.  A companion study by \citet{2017arXiv170607053B} describes an alternative, equally plausible explanation: the BH natal spin depends on the progenitor, such that the most massive BHs are born with low natal spins.  A longer companion study by \citet{DavidePaper} will describe the properties and precessing dynamics of this population in greater detail.

This paper is organized as follows.
First, in Sec. \ref{sec:Popsyn} we describe the entire process used to generate  and characterize detection-weighted populations of precessing binary BHs, evaluated using different assumptions about BH natal kicks.
As described in Sec. \ref{sec:sub:StarTrack}, we adopt previously studied binary evolution calculations to determine how frequently compact binaries merge throughout
the universe.
In Sec. \ref{sec:sub:PN}, we describe how we evolve the binary's precessing BH spins starting from just after it
forms until it enters the LIGO band.
In Sec. \ref{sec:sub:GMM},  we describe the parameters we use to characterize each binary: the component masses and
spins, evaluated after evolving the BH binary according to the process described in Sec. \ref{sec:sub:PN}.  To enable direct comparison with observations, we convert from detection-weighted samples---the output of our binary evolution model---to a smoothed approximation, allowing us to draw inferences about the relative likelihood of generic binary parameters.
In Sec. \ref{sec:Compare} we compare these smoothed models for compact binary formation against LIGO's observations to date.
We summarize our conclusions in Sec. \ref{sec:Conclude}.
In Appendix \ref{ap:gmm-approximation} we describe the technique we use to approximate each of our binary evolution simulations.  In Appendix \ref{ap:HierarchicalCalculation}, we provide technical details of the underlying statistical techniques we use to compare these approximations to LIGO observations.  To facilitate exploration of alternative assumptions about natal spins and kicks, we have made publicly available all of the marginalized likelihoods evaluated in this work, as Supplemental material \cite{supplement}.

\section{\label{sec:Popsyn}Estimating the observed population of coalescing binary black holes}

\subsection{Forming compact binaries over cosmic time}
\label{sec:sub:StarTrack}

\begin{table}
\begin{tabular}{l|r|l|l}
 Name & $\sigma$ (km/s) & $D_{KL}(M)$  & $D_{KL}(m_1,m_2)$ \\ \hline
 M10 & $\O$  &  0.02  & 0.21\\
 M18 & 25 &  0.006 & 0.094 \\
 M17 & 50  & 0 & 0 \\
 M16 & 70  & 0.016 & 0.28  \\
 M15 & 130 & 0.1 & 1.26\\
 M14 & 200 &  0.17 & 1.56\\
 M13 & 265 &  0.40 & 2.1 
\end{tabular}
\caption{Properties of the formation scenarios adopted in this work.  The first column indicates the model calculation
  name, using the convention of  other work \cite{2016Natur.534..512B,2016A&A...594A..97B}.  The second column provides the kick distribution width.  Model M10 adopts
  mass-dependent,  fallback suppressed BH natal kicks.  For the BH population examined here, these natal kicks are
  effectively zero for massive BHs; see, e.g., \cite{2016ApJ...832L...2R}.   The remaining scenarios
  adopt a mass-independent Maxwellian natal kick distribution characterized by the 1-d velocity dispersion $\sigma$, as described in the text.
  The third column quantifies how much the mass distribution changes as we change $\sigma$.  To be concrete, we compare the (source frame) total mass distributions for the BH-BH binaries LIGO is expected to detect, using a Kullback--Leibler (KL) divergence [Eq. (\ref{eq:def:KL})].  If $p(M|\alpha)$
  denotes the mass distribution for $\alpha=$ M10, M18, M17, $\ldots$, and $\alpha_*$ denotes M17, then the third column
  is the KL divergence $D_{KL} (M,\alpha) = \int dM p(M|\alpha)\ln [p(M|\alpha)/p(M|\alpha_*)]$.
  The fourth column is the KL divergence  using the joint distribution
  of both binary masses: $D_{KL}(m_1,m_2|\alpha) = \int dm_1 dm_2
  p(m_1,m_2|\alpha)\ln [p(m_1,m_2|\alpha)/p(m_1,m_2|\alpha_*)]$.  
 Because M10 adopts fallback-suppressed natal kicks, while the remaining models assume fallback-independent natal kicks,
 we use the special symbol $\O$ to refer to M10 in subsequent plots and figures. 
}
\label{tab:Parameters}
\end{table}

Our binary evolution calculations are performed with the \textsc{StarTrack} isolated binary evolution code
\cite{2002ApJ...572..407B,2008ApJS..174..223B},
with updated calculation of common-envelope physics \cite{2012ApJ...759...52D}, compact remant masses
\cite{2012ApJ...749...91F}, and pair instability
supernovae \cite{2016A&A...594A..97B}.
Using this code, we generate a synthetic universe of (weighted) binaries by Monte Carlo \cite{2013ApJ...779...72D}.
Our calculations account for  the time- and metallicity- dependent star formation history of the
universe,  by using a grid of  \textbf{32} different choices for stellar metallicity.  
As shown in Table \ref{tab:Parameters}, we create synthetic universes using the same assumptions (M10) adopted by
default in previous studies  \cite{2016A&A...594A..97B,2016Natur.534..512B,2017arXiv170607053B}.  Again as in
previous work, we explore a one-parameter family of simulations that adopt different assumptions about BH natal kicks
(M13-M18).  Each new model assumes all BHs receive natal kicks drawn from the same Maxwellian distribution, with
one-dimensional velocity distribution parameterized by $\sigma$ (a quantity which changes from model to model).  %
In the M10 model used for reference, BH kicks are also drawn from a Maxwellian distribution, but suppressed by
the fraction of ejected material that is retained (i.e., does not escape to infinity, instead being accreted by the BH).  Because the progenitors of the most massive BHs do
not, in our calculations, eject significant mass to infinity, the heaviest BHs formed in this
``fallback-suppressed kick'' scenario receive nearly or exactly zero natal kicks.

These synthetic universes consist of weighted BH-BH mergers (indexed by $i$), each one acting a proxy for a part of the overall merger rate density in its local volume
\cite{2015ApJ...806..263D,2016ApJ...819..108B}.  As our synthetic universe resamples from the same set of \textbf{32} choices for stellar metallicity, the same evolutionary trajectory appears many times, each at different redshifts and reflecting the relative probability of star formation at different times.  The most frequent formation scenarios and the fraction of detected binaries from each channel are shown in Table \ref{tab:table2}.

The underlying binary evolution calculations performed by \textsc{StarTrack} effectively do not depend on BH spins at any
stage.\footnote{The response of the BH's  mass and spin to accretion depends on the BH's spin.  We adopt a standard procedure whereby the BH accretes from the innermost stable circular
  orbit.  In our binary evolution code, this spin evolution is implemented directly via an ODE
  based on (prograde, aligned) ISCO accretion as in \cite{2000NewA....5..191B}, though the general solution is provided in
  \cite{1970Natur.226...64B}  and applied since, e.g., in 
 \cite{1974ApJ...191..507T,2005ApJ...632.1035O}.  For the purposes of calculating the final BH mass from the natal mass
 and its accretion history, we adopted a BH natal spin of
  $\chi=0.5$; however,  relatively little mass is accreted and  the choice of spin has a highly subdominant
 effect on the BH's evolution.  
} 
 We therefore have the freedom to reuse each calculation above with any BH
natal spin prescription whatsoever.  
Unlike \citet{2017arXiv170607053B}, we do not adopt a physically-motivated and mass-dependent BH natal spin, to
allow us to explore all of the possibilities that nature might allow.    Instead, we treat the
birth spin for each BH as a parameter, assigning spins ${\mathbf \chi}_1$ and $\mathbf{\chi}_2$ to each black
hole at birth.   
For simplicity and without loss of generality, for each event we assume a fixed BH spin for  the first-born 
 ($\chi_1=\left|{\mathbf S}_1\right|/m_1^2$) and a potentially
 different spin for the second-born ($\chi_2 = \left|{\mathbf S}_2\right|/m_2^2$) BH.   Both choices of fixed spin are
parameters.   By carrying out our calculations on a discrete grid in $\chi_1,\chi_2$ for each event---here, we use
$\chi_{1,2} =0.1 \ldots 1$---we encompass a wide
range of possible choices for progenitor spins, allowing us to explore arbitrary (discrete) natal spin
distributions. For comparison, \cite{2017MNRAS.471.2801S} adopted a fixed natal spin $\chi_i=0.7$ for all BHs.
Our choices for BH natal spin distributions are restricted only by our
choice of discrete spins. Our model is
  also implicitly limited by requiring all BHs have natal spins drawn from the same mass-independent distributions.  By design, our
  calculation did not include enough degrees of freedom to enable the natal spin distribution to change with
  mass, as was done for example in \cite{2017arXiv170607053B}.
We assume the progenitor stellar binary is comprised of stars whose spin axes are aligned with the orbital angular
momentum, reflecting natal or tidal \cite{1981A&A....99..126H,2008EAS....29....1M} alignment  (but cf.  \cite{2013ApJ...767...32A}).  %
After the first supernova, several processes could realign the stellar or BH spin with the orbital plane,
including mass accretion onto the BH and tidal dissipation in the star. 
Following \citet{2013PhRvD..87j4028G}, we consider two possibilities.  In our default scenario 
(``no tides''),
spin-orbit alignment is only influenced by BH natal kicks.   In the other scenario (``tides''), tidal dissipation
will cause the stellar spin in stellar-BH binaries to align parallel to the orbital plane.  In the ``tides''
scenario, the second-born stellar spin is aligned with the orbital angular momentum prior to the second SN.    
Following \cite{2013PhRvD..87j4028G}, the ``tides'' scenario assumes  alignment always occurs for merging BH-BH
binaries, independent of the specific evolutionary trajectory involved (e.g., binary separation); cf. the discussion in
\cite{2017arXiv170607053B}. 
In both formation scenarios, we do not allow mass accretion onto the BH to change the BH's spin direction.  Given the  extremely small amount of mass accreted during either conventional or common-envelope
mass transfer, even disk warps and the Bardeen-Petterson effect should not allow the BH spin direction to evolve
\cite{2007ApJ...662..504B,2008ApJ...682..474B,1999MNRAS.305..654K,1998ApJ...506L..97N}.
For coalescing BH-BH binaries the second SN often occurs when the binary is in a tight orbit, with high orbital speed,
and thus less  effect on spin-orbit misalignment \cite{2000ApJ...541..319K,2017PhRvL.119a1101O}.  Therefore, in the ``tides'' scenario, the second-born BH's spin is more
frequently nearly aligned with the final orbital plane, even for large BH natal kicks.  

\begin{table}
\begin{tabular}{l|r}
Formation mechanism & Fraction \\ \hline
MT1(2-1) MT1(4-1) SN1 CE2(14-4;14-7) SN2      &0.261 \\
MT1(4-4) CE2(7-4;7-7) SN1 SN2                  &0.234  \\
MT1(4-1) SN1 CE2(14-4;14-7) SN2                &0.140 \\
MT1(2-1) SN1 CE2(14-4;14-7) SN2                &0.075 \\
MT1(4-4) CE2(4-4;7-7) SN1 SN2                  &0.071 \\
MT1(2-1) SN1 MT2(14-2) SN2                     &0.037 \\
CE1(4-1;7-1) SN1 MT2(14-2) SN2                 &0.028 \\
CE1(4-1;7-1) SN1 CE2(14-4;14-7) SN2            &0.020 \\
CE1(4-1;7-1) CE2(7-4;7-7) SN1 SN2              &0.014\\
MT1(4-4) CE12(4-4;7-7) SN2 SN1                 &0.014 \\
SN1 CE2(14-4;14-7) SN2                         &0.014\\
Other channels  & 0.16
\end{tabular}                
\caption{The most significant formation scenarios and fraction of detected binaries formed from that channel, for the M15 model.  While many of the
  coalescing BH-BH  binaries form via a BH-star binary undergoing some form of stellar mass transfer or interaction, a
  significant fraction of binaries form without any Roche lobe overflow mass transfer after the first SN.  In this example, in the second
  channel alone more than \textbf{23\%} of binaries form without interaction after the first SN.  (The remaining
  formation channels account for \textbf{16\%} of the probability.)  In this notation, integers in braces characterize
  the types of the stellar system in the binary; the prefix refers to different phases of stellar interaction (e.g.,
  MT denotes ``mass transfer,''  SN denotes ``supernova,'' and CE denotes ``common envelope evolution''); and the last
  integer SN$x$ indicates whether the initial primary star  (1) or initial secondary star (2) has
  collapsed and/or exploded to form a BH.   [Some of our BHs are formed without luminous explosions; we use SN to
    denote the death of a massive star and the formation of a compact object.]  A detailed description of these formation channels and
  stellar types    notation is provided in \cite{2002ApJ...572..407B,2008ApJS..174..223B}; in this shorthand, 
 $1$ denotes a main sequence star; 
$2$ denotes  a   Hertzprung gap star; 
$4$ denotes  a core helium burning star;
 $7$ denotes  a main sequence naked helium star; and 
$14$ denotes  a black hole.
 }
\label{tab:table2}
\end{table}

\subsection{Evolving from birth until merger}
\label{sec:sub:PN}
The procedure above produces a synthetic universe of binary BHs, providing binary masses, spins, and orbits just after
the second BH is born.   Millions to billions of years must pass before these binaries coalesce, during which time the
orbital and BH spin angular momenta precess substantially
\cite{2015PhRvL.114h1103K,2015PhRvD..92f4016G}.  
We use precession-averaged 2PN precessional dynamics, as implemented in \textsc{precession} \cite{2016PhRvD..93l4066G}, to evolve the spins from birth until the binary BH orbital
frequency is 10Hz (i.e., until the GW frequency is 20Hz); see \cite{DavidePaper} for details. 
When identifying initial conditions, we assume the binary has already efficiently circularized.
When identifying the final separation,  we only use the
Newtonian-order relationship between separation and orbital frequency. 
The \textsc{precession}  code is publicly available at \href{https://github.com/dgerosa/precession}{github.com/dgerosa/precession}. %

\subsection{Characterizing the observed distribution of binaries}
\label{sec:sub:GMM}

\begin{figure}
\includegraphics[width=\columnwidth]{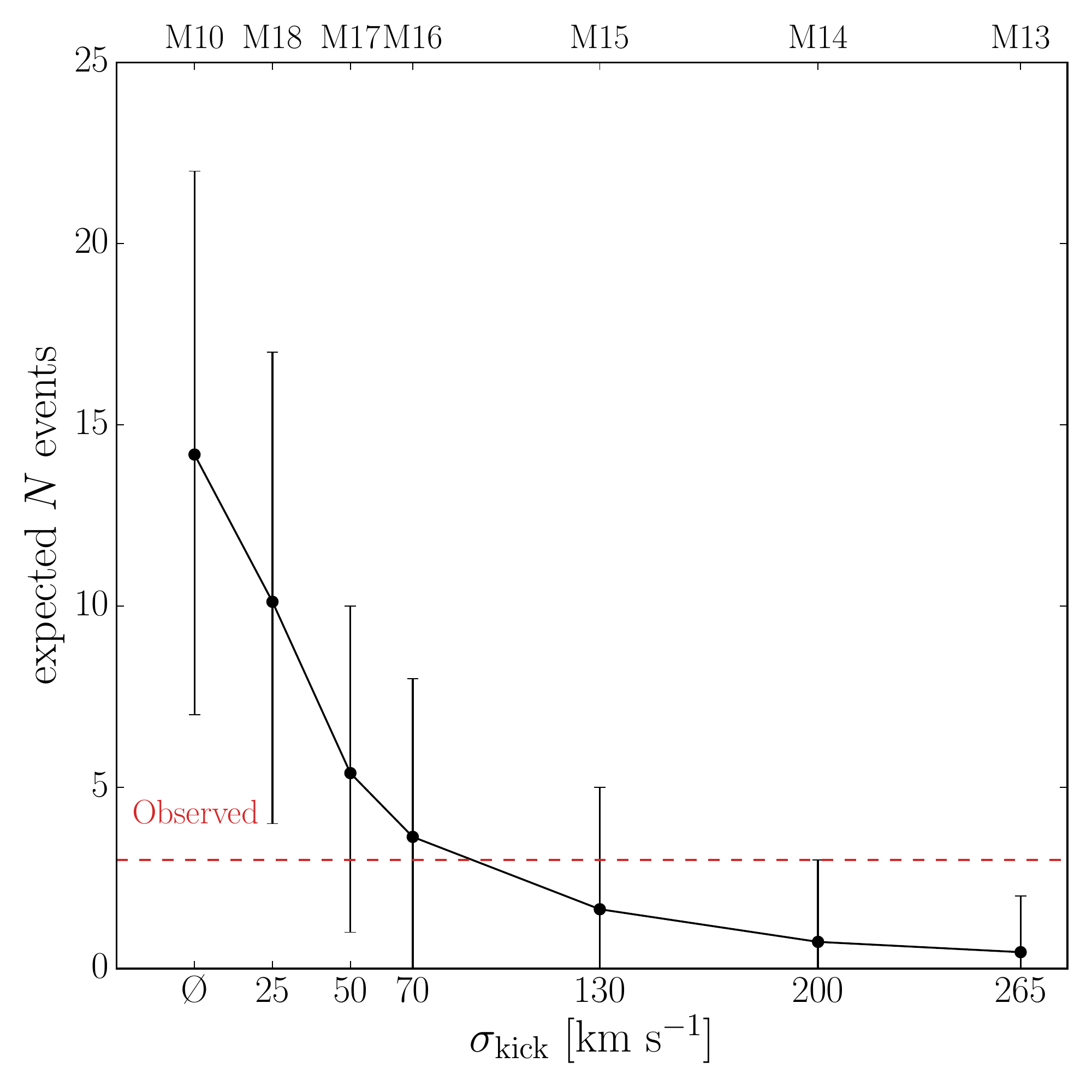}
\caption{
  \label{fig:DetectionNumber}
  \emph{Expected number of events versus kick strength}:
  Expected number of BH merger detections predicted at LIGO's O1 sensitivity and for the duration of O1
  by our formation scenarios.   The predicted number of events decreases
  rapidly as a function of the BH natal kick.  Also shown is the 95\% confidence
  interval, assuming Poisson distribution with mean predicted by our
  model.  This purely statistical error bar does not account for any model systematics (e.g., in the overall star
  formation rate and metallicity history of the universe).  
  The horizontal red dashed line corresponds to the number (3) of observations reported in O1 \cite{2016PhRvX...6d1015A}.
}
\end{figure}

At the fiducial reference frequency adopted in this work
($20\unit{Hz}$), 
a binary BH is characterized by its
component masses and its (instantaneous) BH spins $\mathbf{S}_{1,2}$.   For the heavy BHs of 
current interest to LIGO, the principal effect of BH spin on the
orbit and emitted radiation occurs through the spin combination
\begin{align}
\chi_{\rm eff} &= (\mathbf{S}_1/m_1 + \mathbf{S}_2/m_2) \cdot \mathbf{\hat{L}}/(m_1+m_2) \nonumber \\
 & = (\chi_1 m_1 \cos \theta_1 + \chi_2 m_2 \cos \theta_2)/(m_1+m_2),
\end{align}
where $\theta_{1,2}$ denote the angles between the orbital angular momentum and the component BH spins.  
That said, depending on the duration and complexity of the source responsible, GW measurements  may also provide  additional  constraints on the underlying spin directions themselves
\cite{2016PhRvD..93d4071T}, including on the spin-orbit misalignment angles $\theta_{1,2}$. 
For the purposes of this work, we will be interested in the (source-frame) binary masses $m_1,m_2$ and the spin
parameters $\chi_{\rm eff},\theta_1,\theta_2$, as an approximate
characterization of the most observationally accessible degrees of freedom;
cf. \citet{2017MNRAS.471.2801S}, which used $\theta_{1,2}$, and  
\citet{2016PhRvD..93d4071T}, which used $\theta_{1,2}$ and 
the angle $\Delta\Phi$ between the spins' projection
onto the orbital plane. In particular, $\Delta\Phi$ is well-known to contain valuable information
\cite{2013PhRvD..87j4028G} and be observationally accessible \cite{2016PhRvD..93d4071T}.  At present, the preferred model adopted for parameter inference, known as IMRPhenomP, does not incorporate
the necessary degree of freedom \cite{2014PhRvL.113o1101H}, so we cannot  incorporate its effect here.  With additional
and more informative binary black hole observations, however, our method should be extended to employ all of the spin degrees of freedom,
particularly $\Delta\Phi$.   As input, this extension will require  inference results that incorporate the
effect of two  two precessing spins, either by using semianalytical models
\cite{2014PhRvD..89f1502T,2017PhRvD..96b4058B,2017PhRvD..95j4023B} or by using numerical relativity \cite{2016PhRvD..94f4035A}.  
We adopt a conventional model for LIGO's sensitivity to a population of binary BHs
\cite{2010ApJ...716..615O,2015ApJ...806..263D,2016ApJ...833L...1A}. 
In this approach,  LIGO's sensitivity is limited by the second-most-sensitive interferometer, using a detection
threshold signal-to-noise ratio $\rho=8$ and the fiducial detector
sensitivity reported for O1 \cite{2016PhRvX...6d1015A}.  This sensitivity model is a good approximation to the performance
reported for both in O1 and early in  O2 \cite{2017PhRvL.118v1101A}.  Following
\cite{2015ApJ...806..263D,2016ApJ...819..108B}, we use the quantity
$r_i$ [Eq.~(8) in
  \cite{2016ApJ...819..108B}] to  account for
the contribution of this binary to LIGO's detection rate  in our
synthetic universe, accounting for the size of the universe at the time the binary coalesces and LIGO's orientation-dependent
sensitivity.     For simplicity and following previous work \cite{2015ApJ...806..263D,2016ApJ...833L...1A}, we estimate the
detection probability without accounting for the effects of BH spin.
Previous studies have used this detection-weighted procedure to evaluate and report on the expected 
distribution of binary BHs detected by LIGO
\cite{2016Natur.534..512B,2016A&A...594A..97B,2017arXiv170607053B}.
Since the same binary evolution  $A$ occurs many times in  our synthetic universe, we simplify our results by computing
one  overall detection rate $r_A =
\sum_{i\in A} r_i$ for each evolution.  
When this procedure is performed,  relatively few distinct binary evolutions $A$ have significant weight.  
 While our synthetic universe contains millions of binaries,
only $O(10^4)$  distinct BH-BH binaries are significant in our final results for each of the formation scenarios listed
in Table \ref{tab:Parameters}. 
Fig. \ref{fig:DetectionNumber} illustrates the expected detected number versus assumed BH natal kick strength.

The significant BH natal kicks adopted in all of our formation scenarios (except M10) frequently produce significant
spin-orbit misaligment. Fig. \ref{fig:MisalignmentExample} shows that strong misalignment occurs ubiquitously, even
for small BH natal kicks; see \cite{DavidePaper} for more details.  %
This strong spin-orbit misalignment distribution produces an array of
observationally accessible signatures,
most notably via an invariably wide distribution of $\chi_{\rm eff}$.
In \cite{DavidePaper} the distribution was constructed for all of our models, finding that (except for M10)
considerable support exists  for $\chi_{\rm eff}<0$. %
Our calculation is fully consistent with the limited initial exploration reported in \citet{2016ApJ...832L...2R}, which claimed $\chi_{\rm eff}<0$ was implausible except for extremely large natal kicks.  Their collection of calculations explored fallback-suppressed kicks (e.g., equivalent to our model M10);  adopted natal kicks larger than we explored here; or adopted  mass-dependent natal kicks.  We show that significant spin-orbit misalignment is plausible if all BHs---even massive ones---receive a modest natal kick. %
BH natal kicks therefore provide a robust  mechanism to explain the observed  $\chi_{\rm eff}$ and spin-orbit misalignments reported by LIGO for its first few detections.

\begin{figure}
\includegraphics[width=\columnwidth]{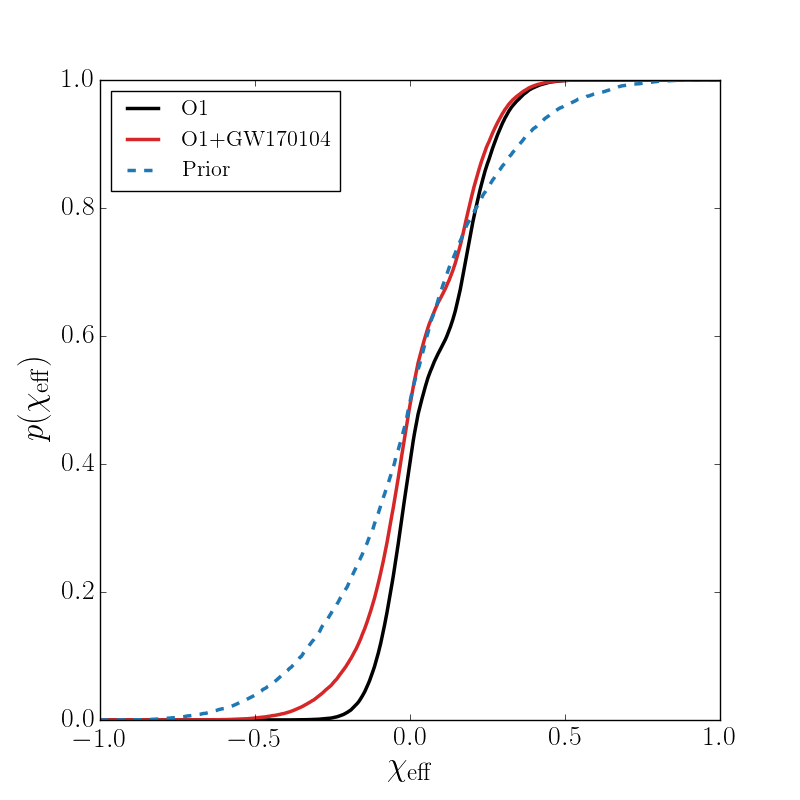}
\caption{\label{fig:ChiEffEmpiricalCDF}\emph{Empirical cumulative distribution function for $\chi_{\rm eff}$}:  The solid blue line shows
  the conventional  prior distribution for $\chi_{\rm eff}$, generated by selecting masses uniformly in $m_{1,2} \ge 1 M_\odot$, $m_{1,2}\le
  100 M_\odot$,  $m_1+m_2<100 M_\odot$, and isotropic spins generated independently and uniformly in magnitude.  This
  prior was adopted when analyzing all LIGO events.  The
  solid black line shows the empirical cumulative distribution for $\chi_{\rm eff}$,   derived from the collection of
  events $\alpha=$ GW150914, GW151226, and LVT151012 via their 
 posterior cumulative  distributions $P_\alpha(\chi_{\rm eff})$ via $P(\chi_{\rm eff}) = \sum_\alpha P_\alpha(\chi_{\rm
   eff})/3$.  In this curve, the posterior distributions are provided by LIGO's full O1 analysis results \cite{2016PhRvX...6d1015A}, as described
 in the text.  The solid red
 line shows the corresponding result when GW170104 is included.  The approximate posterior distribution for GW170104 is
 based on published results, as described in Appendix \ref{ap:mockup-GW170104}.  
}
\end{figure}

The procedure described above \emph{samples} a synthetic universe and synthetic observations by LIGO.  However, to
compare to LIGO's observations, we need to be able to assess the likelihood of \emph{generic} binaries according to our
formation scenario, extrapolating between what we have  simulated.  We therefore
estimate the merger rate distribution as a function of binary masses, spins, and spin-orbit misalignments.  Our estimate
uses a carefully calibrated Gaussian mixture model, with special tuning as needed to replicate sharp features in our mass and
misalignment distribution; see Appendix \ref{ap:gmm-approximation} for details.   

\section{Comparison with gravitational wave observations}
\label{sec:Compare}

\subsection{Gravitational wave observations of binary black holes}

During its first observing run of $T_1=\TdaysOone \; \unit{days}$,
LIGO has reported the observation of three BH-BH mergers: GW150914,
LVT151012, and GW151226
\cite{2016PhRvL.116f1102A,2016PhRvL.116x1103A,2016PhRvX...6d1015A}.  In an analysis
of $T_2=\TdaysAnalyzedOtwoEvent~\unit{days}$ of data from its second
observing run, at comparable sensitivity, LIGO has since reported the
observation of another binary BH: GW170104
\cite{2017PhRvL.118v1101A}.
To draw out more insight from each observation, rather than use the
coarse summary statistics LIGO provides in tabular form, %
we employ the
underlying posterior parameter distribution estimates provided by the
LIGO Scientific Collaboration for the three O1 events
\cite{2016PhRvX...6d1015A,2016PhRvL.116x1102A,2015PhRvD..91d2003V}.  For GW170104, we instead adopt an approximate posterior
distribution  described in
Appendix \ref{ap:mockup-GW170104} based solely on reported tabular results; that said, we
are confident that this approximation makes no difference to our conclusions. 
For each event, for brevity indexed by an integer $n=1,2,3,\ldots,N$, these estimates are generated by comparing 
a proposed gravitational wave source $x$
with the corresponding stretch of gravitational wave data $d$ using a (Gaussian) likelihood function $p(d|x)$ that
 accounts for the frequency-dependent sensitivity of the detector  (see, e.g., \cite{2015PhRvD..91d2003V,2016PhRvL.116x1102A,2016PhRvD..94f4035A} and
references therein).  In this
expression $x$ is shorthand for the 15 parameters needed to fully specify a quasicircular BH-BH binary in space and
time, relative to our instrument; and $d$ denotes all the gravitational wave data from all of LIGO's instruments.
This analysis adopts prior assumptions about the relative likelihood of different progenitor binaries $p_{\rm ref}(x)$:  equally likely to have any pair of component masses, any spin direction, any spin magnitude, any
orientation, and any point in spacetime (i.e., uniform in comoving volume).   Then, using standard Bayesian tools
\cite{2016PhRvL.116x1102A,2015PhRvD..91d2003V}, the LIGO analysis produced 
 a sequence of independent, identically distributed samples
$x_{n,s}$ ($s=1,2,\ldots,S$) from the posterior distribution for each event $n$; that is, each $x_{n,s}$ is
drawn from a distribution proportional to $p(d_n|x_n)p_{\rm ref}(x_n)$.
This approach  captures degeneracies in the posterior not
previously elaborated in detail, most notably the well-known strong correlations between the inferred binary's component masses
and spins (e.g., between $\chi_{\rm eff}$ and $m_2/m_1$).
\footnote{Different properties of the binary, like the masses and spins, influence the inspiral, and thus the radiation $h(t)$, in generally different ways; however, sometimes, several parameters can influence the radiation in a similar or degenerate way.  For example, both the binary mass ratio and (aligned) binary spin can extend the duration of the inspiral.  Similarly, both the binary masses and spins---8 parameters---determine the final complex frequency of the BH---at leading order, only set by two parameters.
Due in part to degeneracies like these, LIGO's inferences about the parameters $x$ for each merging BH lead to a highly correlated likelihood $p(d|x)$ and hence posterior distribution; see, e.g. \cite{2015PhRvD..91d2003V,2016PhRvL.116x1102A,2016PhRvD..94f4035A} and references therein.}
Equivalently, this approach gives us direct access to properties of the posterior distribution that were not reported
  in published tables \cite{2016PhRvX...6d1015A}, most notably for the relative posterior probabilities of different choices for
  binary BH spins (e.g., the data underlying Fig. \ref{fig:ChiEffEmpiricalCDF}).

\subsection{Comparing models to observations}
\label{sec:sub:CompareMarginalLikelihood}
The overall likelihood of GW data $\{d\}$ using a model parametrized by $\Lambda$ is \cite{2004AIPC..735..195L}
\begin{eqnarray}
\label{eq:Likelihood}
  p(\{d\}|\Lambda)
  \propto
  e^{-\mu}
  \prod_n
    \int \mathrm{d}x_n \; p(d_n| x_n) \; \mathcal{R} \, p(x_n|\Lambda)
\end{eqnarray}
where $x_n$ denote candidate intrinsic and extrinsic parameters for the $n$th observation, $\mu$ is the expected number
of detections according to the formation scenario $\Lambda$, $p(d_n|x_n)$ is the likelihood for event $n$;
$p(\{d\}|\Lambda)$ is the marginalized likelihood; $p(x_n|\Lambda)$ is the prior evaluated at event $n$; and $\mathcal{R}$ (implicitly depending on $\Lambda$ as
well) is the average number of merger events
per unit time and volume in the Universe.  In this expression, we have subdivided the data $\{d\}$ into data with
confident detections $d_1,d_2, \ldots, d_N$ and the remaining data; the Poisson prefactor $\exp(-\mu)$ accounts for the
absence of detections in the remaining data; and the last product accounts for each independent observation $d_n$.   
Combined, the factors $ e^{-\mu} \prod_n \mathcal{R} p(x_n)$ are the  distribution function for an inhomogeneous Poisson
process used to
characterize the formation and detection of coalescing BH binaries \cite{2015ApJ...810...58S,2013PhRvD..88h4061O}.   
As described in Appendix
\ref{ap:HierarchicalCalculation},  the probability density functions $p(x|\Lambda)$ are estimated from the weighted
samples that define each synthetic universe $\Lambda$, and the integrals $\int p(d|x)p(x|\Lambda)$ are performed efficiently via Monte Carlo
integration.
Similarly, the expected number of detections $\mu $  at O1 sensitivity---a known constant for each model $\Lambda$---is already provided by the detailed cosmological
integration performed in prior work; see Sec. \ref{sec:Popsyn} and Fig. \ref{fig:DetectionNumber}.
Since the marginal likelihood can always be evaluated, the  model inference on our discrete set of models becomes an application of  Bayesian statistics.  In this work, we report the Bayes factor or likelihood ratio $K_{ij} = p(\{d\}|\Lambda_i)/p(\{d\}|\Lambda_j)$ between two
different sets of assumptions.   
To fix the zero point for the log Bayes factor, we adopt the M16 model
with $\chi_1=\chi_2=0.5$, 
henceforth denoted
collectively as $J$, and henceforth use  $\ln K$ as shorthand for $\ln K_{iJ}$.  %
In what follows, we will mainly discuss comparisons of our models to  all of LIGO's reported detection candidates in O1: GW150914, GW151226, and
LVT151012 \cite{2016PhRvX...6d1015A}.
 We do this because LIGO's
O1 observational time and survey results are well-defined and comprehensively reported \cite{2016PhRvX...6d1015A}; because we can employ detailed
inference results for all O1 events; and because, as we show below, adding GW170104 to our analysis produces little
change to our results.    Using the approximate posterior described in Appendix \ref{ap:mockup-GW170104} for GW170104, we will also compare all reported LIGO observations (O1 and GW170104) to our models.
Critically, for clarity and to emphasize the information content of the data, in several of our figures we will illustrate the marginal
likelihood of the data $p(\{d\}|\Lambda)$ evaluated assuming all binaries are formed with \emph{identical} natal spins.  These
strong assumptions in our illustrations show just how much the data informs our understanding of BH natal spins.   With only four observations,
assumptions about the spin distribution are critical to make progress.   As described in Appendix
\ref{ap:HierarchicalCalculation}, we can alternatively evaluate the marginal likelihood accounting for any concrete spin
distribution,  or even all possible spin
distributions---in our context,  all possible mixture combinations of the $100$ different choices for $\chi_1$ and $\chi_2$
that we explored.  In the latter case, as we show   below, just as one expects \emph{a priori}, observations cannot significantly inform this
100-dimensional posterior spin distribution.    As suggested in previous studies \citep[e.g.][]{2016ApJ...818L..22A,2016PhRvD..94f4035A,2017Natur.548..426F,2017arXiv170607053B}, LIGO's
observations in O1 and  O2 can be fit by  models that includes a wide range of
progenitor spins, so long as sufficient probability exists for small natal spin and/or significant misalignment.  
As a balance between complete generality on the one hand (a 100-dimensional distribution of natal spin distributions)
and implausibly rigid assumptions on the other (fixed natal spins), we emphasize a simple one-parameter model, where BH
natal spins $\chi$ are drawn from the piecewise constant distribution
\begin{eqnarray}
p(\chi) = \begin{cases}
 \lambda_A/0.6  &  \chi\le 0.6 \\
(1-\lambda_A)/0.4 & 0.6<\chi<1
\end{cases}
\label{eq:TwoBinSpinModel}
\end{eqnarray}
where $\lambda_A$ is the probability of a natal spin $\le 0.6$ and the choice of cutoff $0.6$ is motivated by our results below.

\section{Results}

In this section we calculate the Bayes factor $\ln K$ for each of the binary evolution models described above.  Unless
otherwise noted, we compare our models to LIGO's O1 observations (i.e., the observation of GW150914, GW151226, and
LVT151012),  using each model's correlated predictions for the event rate, joint mass
distribution ($m_1,m_2$), $\chi_{\rm eff}$ distribution, and the distribution of $\theta_1,\theta_2$.   
For numerical context, a Bayes factor of $\ln 10 \simeq 2.3$ is by definition equivalent to 10:1 odds in favor of some model over
our reference model.   Bayes factors that are more than $5$ below the largest Bayes factor observed are in
effect implausible (e.g., more than 148:1 odds against), whereas anything within $2$ of the peak are reasonably likely. 

\begin{figure*}
\includegraphics[width=\columnwidth]{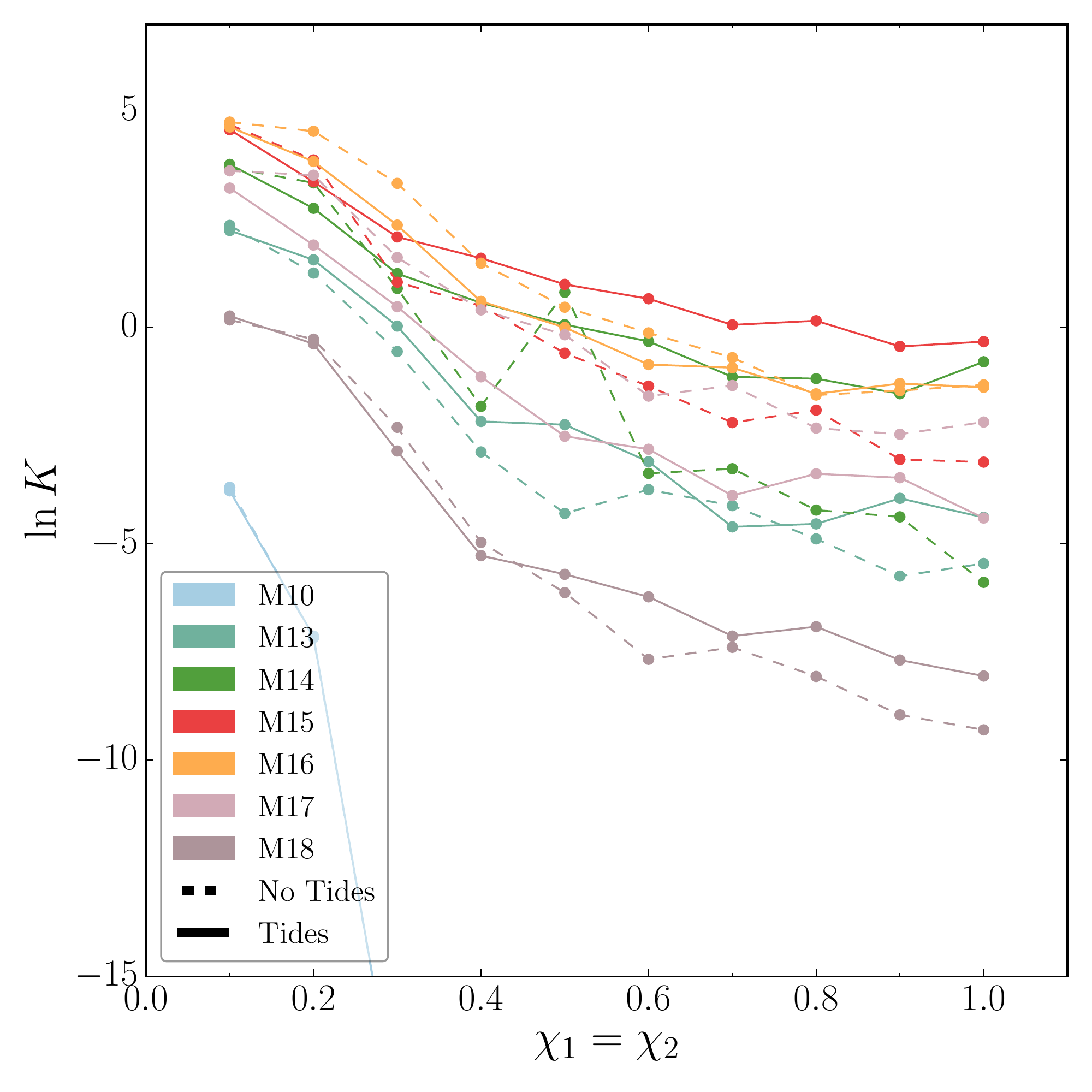}
\includegraphics[width=\columnwidth]{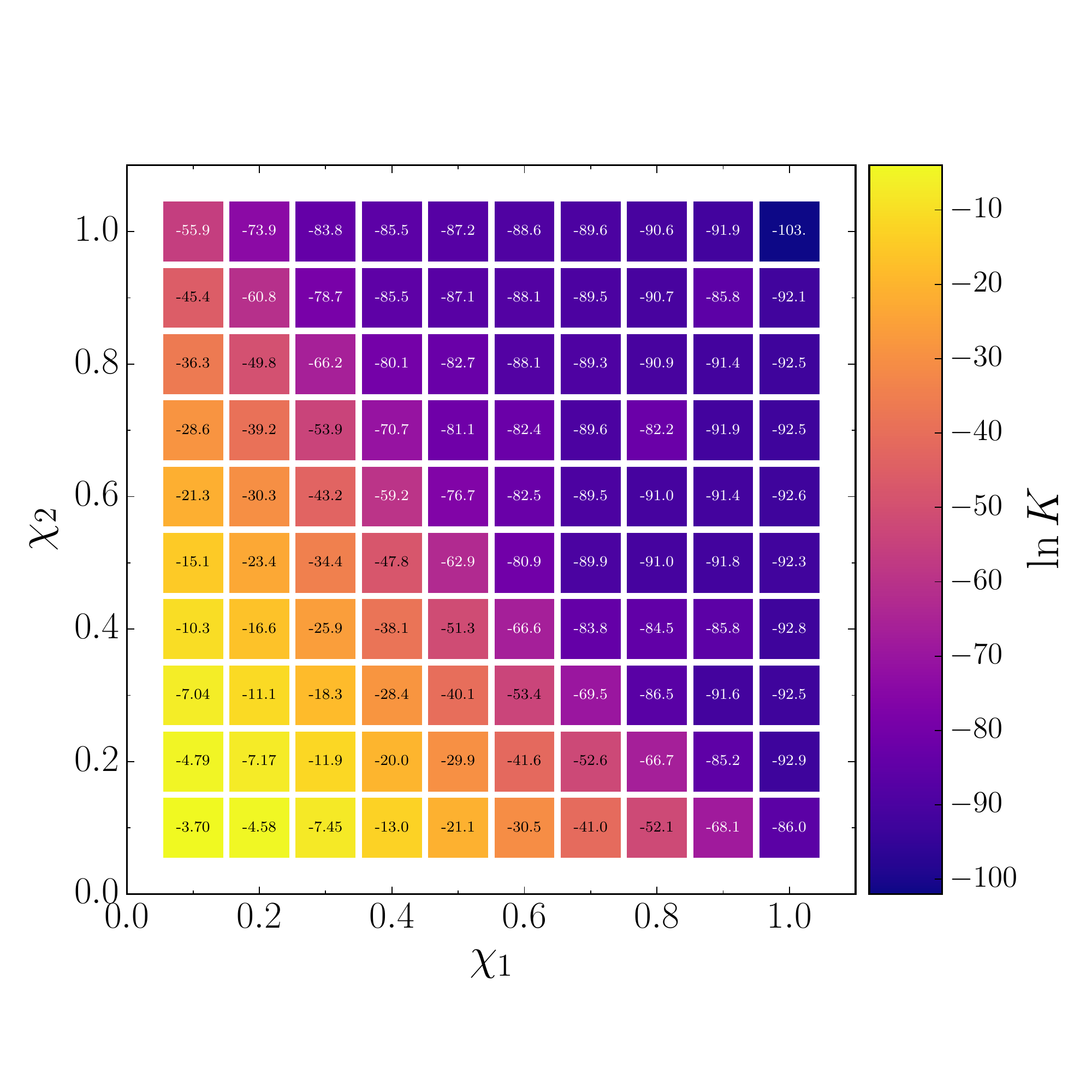}
\caption{\label{fig:Results1:M10:LowSpin}\emph{Standard small-kick scenario (M10) requires small natal BH spin}: Left panel: A plot of the Bayes factor $K$ derived by
  comparing GW151226, GW150914, and LVT151012 to the M10 (blue) formation scenario, versus the magnitude of assumed BH
  natal spin $\chi_1=\chi_2$.  All other models are shown for comparison.  Colors and numbers indicate the
  Bayes factor; dark colors denote particularly unlikely configurations.  Right panel: As before (i.e., for M10), but in two dimensions, allowing the BH natal spins for the
  primary and secondary BH to be independently selected (but fixed); color indicates the Bayes factor.   As this scenario predicts strictly aligned spins for the heaviest BH-BH binaries, only small BH natal spins are consistent with LIGO's
constraints on the (aligned) BH spin parameter $\chi_{\rm eff}$ in O1 (and GW170104); see
  \citet{2016PhRvD..94f4035A,2017PhRvL.118v1101A} and 
  \cite{2017Natur.548..426F}.
}
\end{figure*}

\subsection{Standard scenario and  limits on BH natal spins (O1)}
The M10 model allows us to examine the implications of binary evolution with effectively zero natal kicks.  %
The M10 model adopts fiducial assumptions about binary evolution and BH natal kicks, as described in prior work
\cite{2016Natur.534..512B,2016A&A...594A..97B}.  In this model, BH kicks are suppressed by fallback; as a result, the
heaviest BHs  receive nearly or exactly zero natal kicks and hence have nearly or exactly zero
spin-orbit misalignment.   %

If heavy BH binaries have negligible spin-orbit misalignment, then  natal BH spins are  directly constrained from LIGO's
measurements (e.g., of $\chi_{\rm eff}$).   For example,  LIGO's observations of GW150914 severely constrain 
its component spins to be small, if the spins must be strictly and positively aligned \cite{2016PhRvL.116x1102A,2016PhRvD..94f4035A}.
   Conversely, however, LIGO's observations for GW151226 require some nonzero spin.
Combined, if we assume all BHs have spins drawn from the same, mass-independent distribution and have negligible
spin-orbit misalignments, then we conclude BH natal
spins should be preferentially small.   [We will return to this statement in Sec. \ref{sec:sub:SpinDistributionIssues}.]

Fig. \ref{fig:Results1:M10:LowSpin} shows one way to quantify  this effect within the context of our calculations.
The left panel shows the Bayes factor for all of our formation models (including M10) as a function of BH natal spin,  assuming all BHs have the same (fixed) natal spin
$\chi=\chi_1=\chi_2$.  As expected from LIGO's data, large natal BH spins cannot be adopted with M10 and remain
consistent with LIGO's observations. 
The right panel shows the Bayes factor for M10 as function of both BH natal spins, allowing the more massive and less
massive BHs to receive different (fixed) natal spins.  [The blue line on the left panel uses precisely the same data as
  the diagonal $\chi_1=\chi_2$ on the right.]  The colorscale graphically illustrates the same conclusion: 
 though marginally greater freedom exists for natal   BH spin on the smaller of the two BHs,  we can rule out
 that all BHs, independent of their mass, have significant  natal spin if M10 is true.  Conversely, if M10 is true and
 all BHs have the same natal spins, then this natal spin is likely small.

\subsection{BH natal kicks and misalignment (O1)}

\begin{figure*}
\includegraphics[width=\columnwidth]{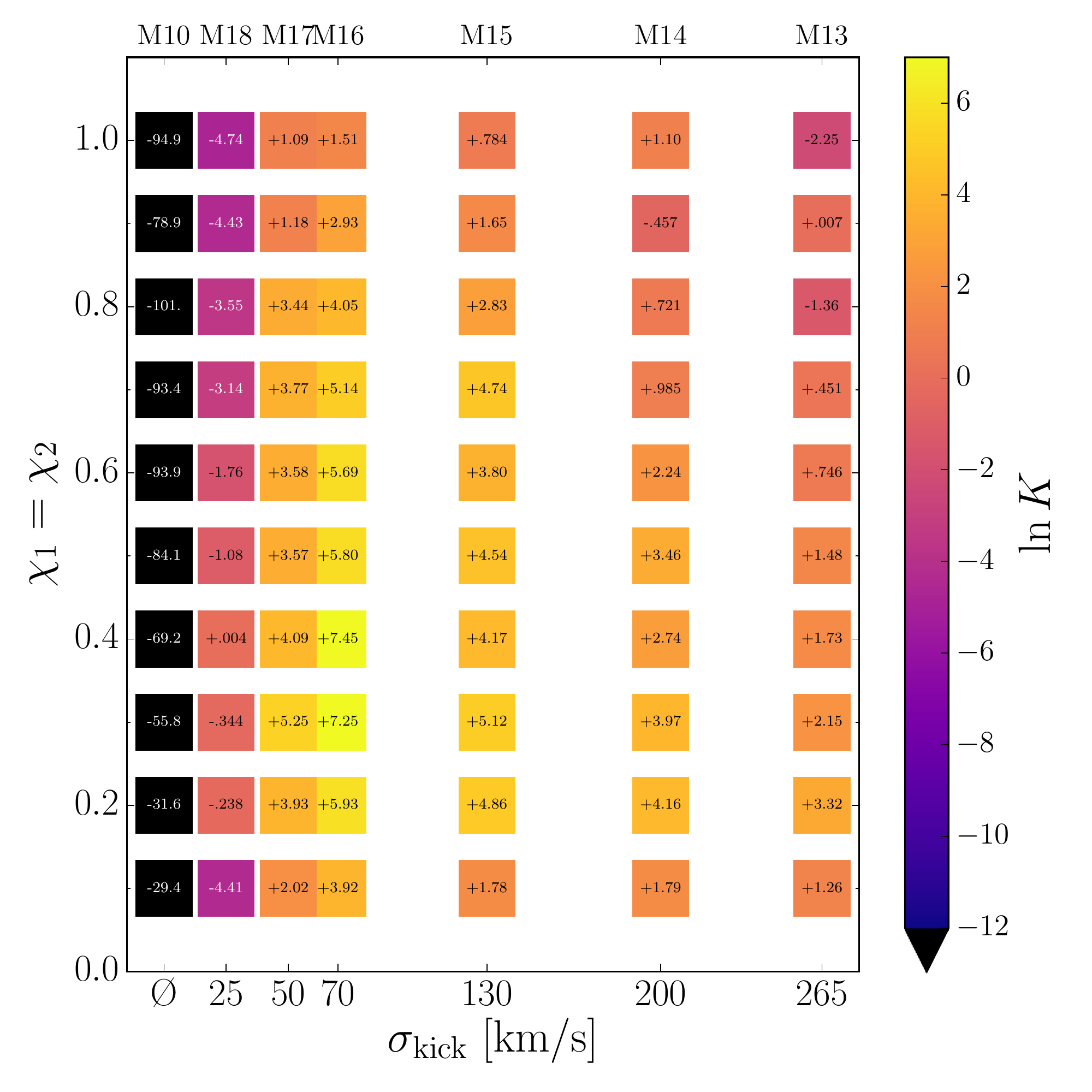}
\includegraphics[width=\columnwidth]{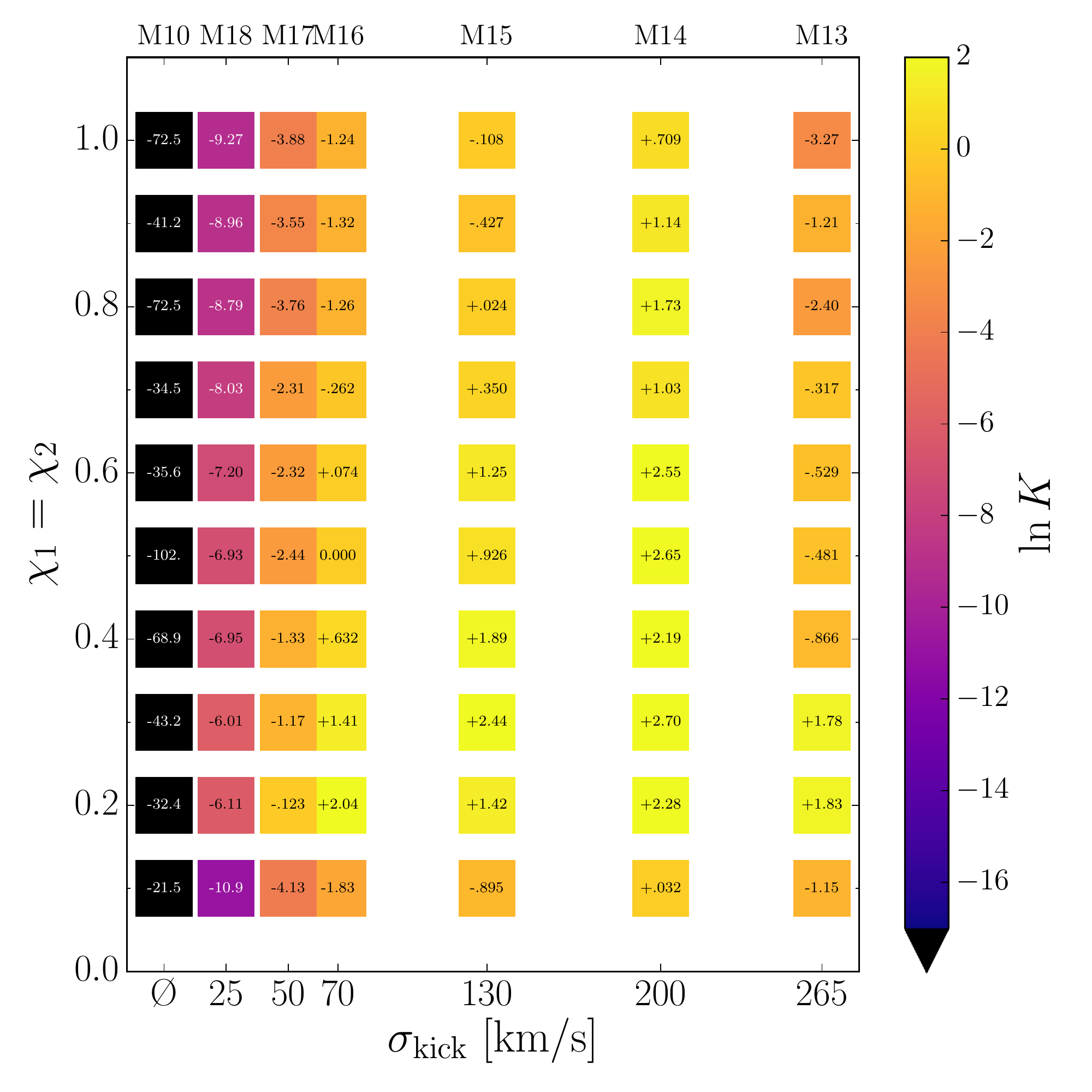}
\includegraphics[width=\columnwidth]{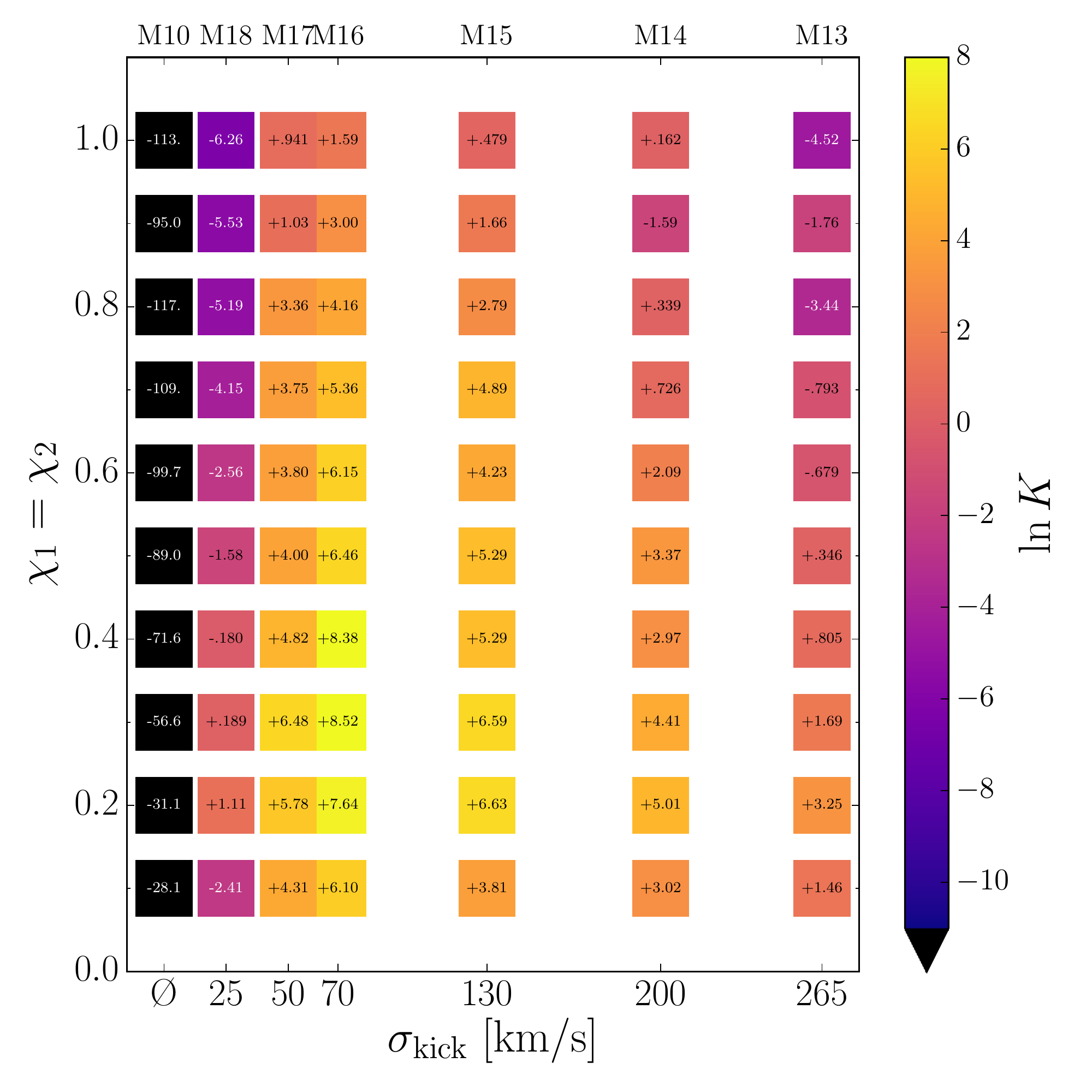}
\includegraphics[width=\columnwidth]{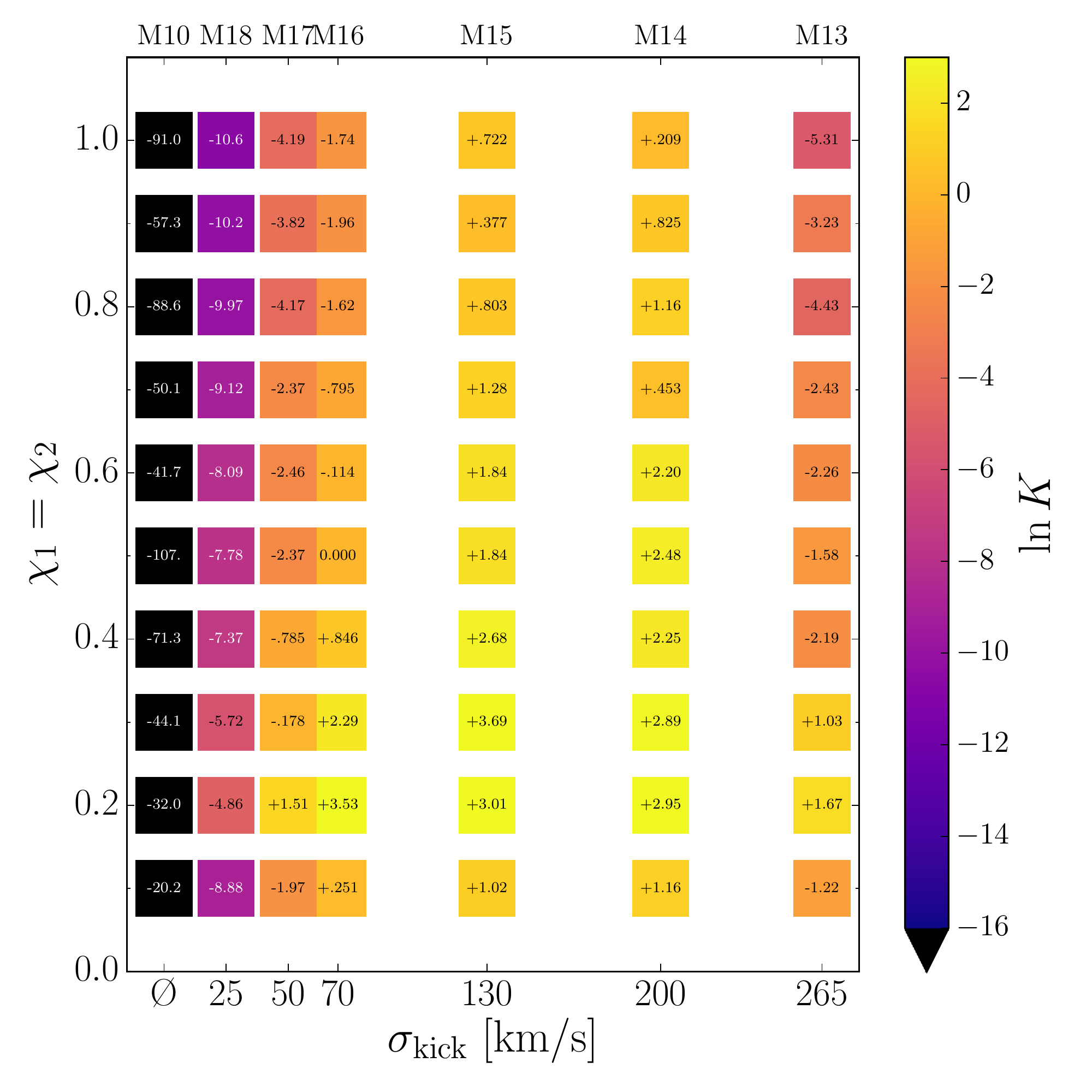}
\caption{\label{fig:Results5:VersusChiKick}\emph{Bayes factor versus spin and kicks, with and without tides}: A plot of
  the Bayes factor versus BH natal spin ($\chi=\chi_{1}=\chi_2$) and natal kick ($\sigma_{\mathrm{kick}}$).  The left and right panels
  correspond to ``no tides'' and ``tides,'' respectively.
The top two panels use only the O1 events; the bottom two panels account for the events and network sensitivity updates
reported in the GW170104 discovery paper.    In each panel, the zero point of the Bayes factor is normalized to the
BH-BH formation scenario with $\chi_1=\chi_2=0.5$ and $\sigma=70\; \unit{km/s}$ and ``tides.''
}
\end{figure*}

In the absence of BH natal kicks, the preponderance of observed BH-BH binaries consistent with $\chi_{\rm eff} \simeq 0$
(e.g.,  GW150914 and GW170104)
provided \emph{conditional} evidence in favor of small BH natal spins.  But even small BH natal kicks can frequently produce
significant spin-orbit misalignment.  Once one incorporates models that permit nonzero BH natal kicks,  then
even binary BHs with large BH natal spins could be easily reconciled with every one of LIGO's observations.   Figures
\ref{fig:Results1:M10:LowSpin} and \ref{fig:Results5:VersusChiKick} %
provide a quantitative illustration of just how much more easily models with even modest
BH natal kicks can explain the data, for a wide range of BH spins.  
When natal kicks greater than $25\unit{km/s}$ are included, the BH natal spin is nearly unconstrained.    As
is particularly apparent in Fig.  \ref{fig:Results5:VersusChiKick}, %
some natal BH spin is required to reproduce the nonzero spin seen in GW151226. 

Larger kicks produce frequent, large spin-orbit misalignments and therefore greater consistency with the properties of
all of LIGO's observed binary BHs.  Spin-orbit misalignment is consistent with
the spin distribution of GW151226, and helpful to explain the distribution of $\chi_{\rm eff}$ for LIGO's other observations.  
 However, larger kicks also disrupt more binaries, substantially decreasing
the overall event rate (see Fig. \ref{fig:DetectionNumber}).   
Fig. \ref{fig:Results5:VersusChiKick}  %
 illustrates the tradeoff between spin-orbit misalignment and event rate.

\subsection{Tides and realignment (O1)}

All other things being equal, our ``no  tides''  scenarios most frequently produce  significant spin-orbit
misalignment.  As a result, even for large BH natal spins, these models have a greater ability to explain LIGO's
observations, which are largely consistent with $\chi_{\rm eff}\simeq 0$.  The ``tides'' scenario produces smaller
misalignments for the second-born BH.
Fig. \ref{fig:Results5:VersusChiKick} %
quantitatively illustrates how the ``no tides'' scenario marginally fits the data
better. %
In order to reproduce the inferred distribution of spin-orbit misalignments (in GW151226) and low $\chi_{\rm eff}$ (for all
events so far), the
``tides'' models likely have   (a) larger BH natal kicks $\simeq 200 \unit{kms/s}$ and (b) low BH natal spins
$\chi_{1,2} \lesssim 0.2$. %
Conversely, when ``no tides'' act to realign the second BH spin, %
 \emph{small} natal kicks  $\simeq 50
\unit{km/s}$ are  favored.  
Fig. \ref{fig:Results5:VersusChiKick} illustrates the two distinct conclusions about BH natal kick strength drawn,
depending on whether stellar tidal realignment is efficient or inefficient.
Based on this figure (and hence on the assumption of fixed natal spins), we estimate that massive BHs should receive a
natal kick of  $\sim $ 50 km/s if no processes act to realign
stellar spins.  Significantly larger natal kicks, with one-dimensional velocity dispersion $\simeq 200\unit{km/s}$,  will be required if stellar spins efficiently
realign prior to the second BH's birth.    

\begin{figure*}
\includegraphics[width=\columnwidth]{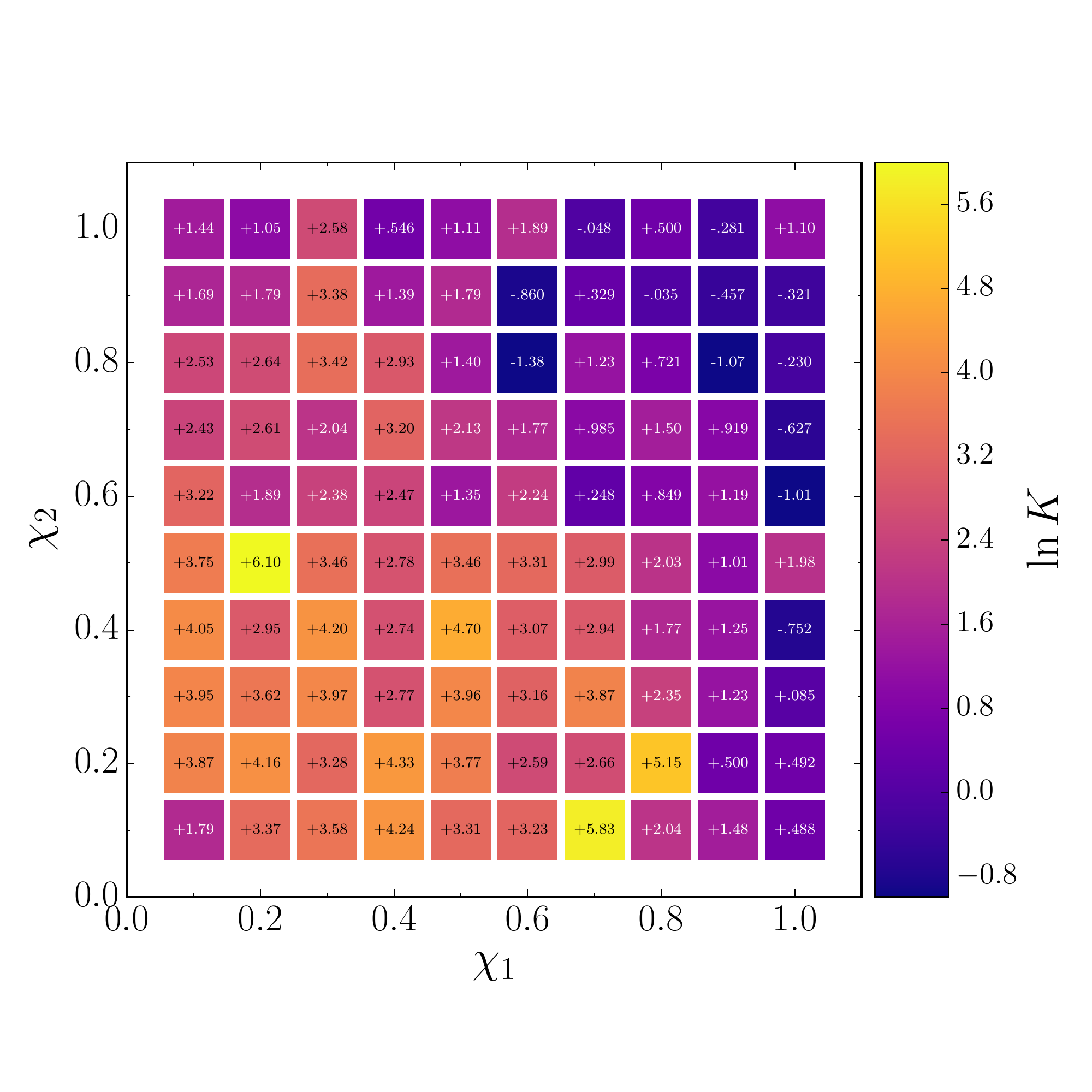}
\includegraphics[width=\columnwidth]{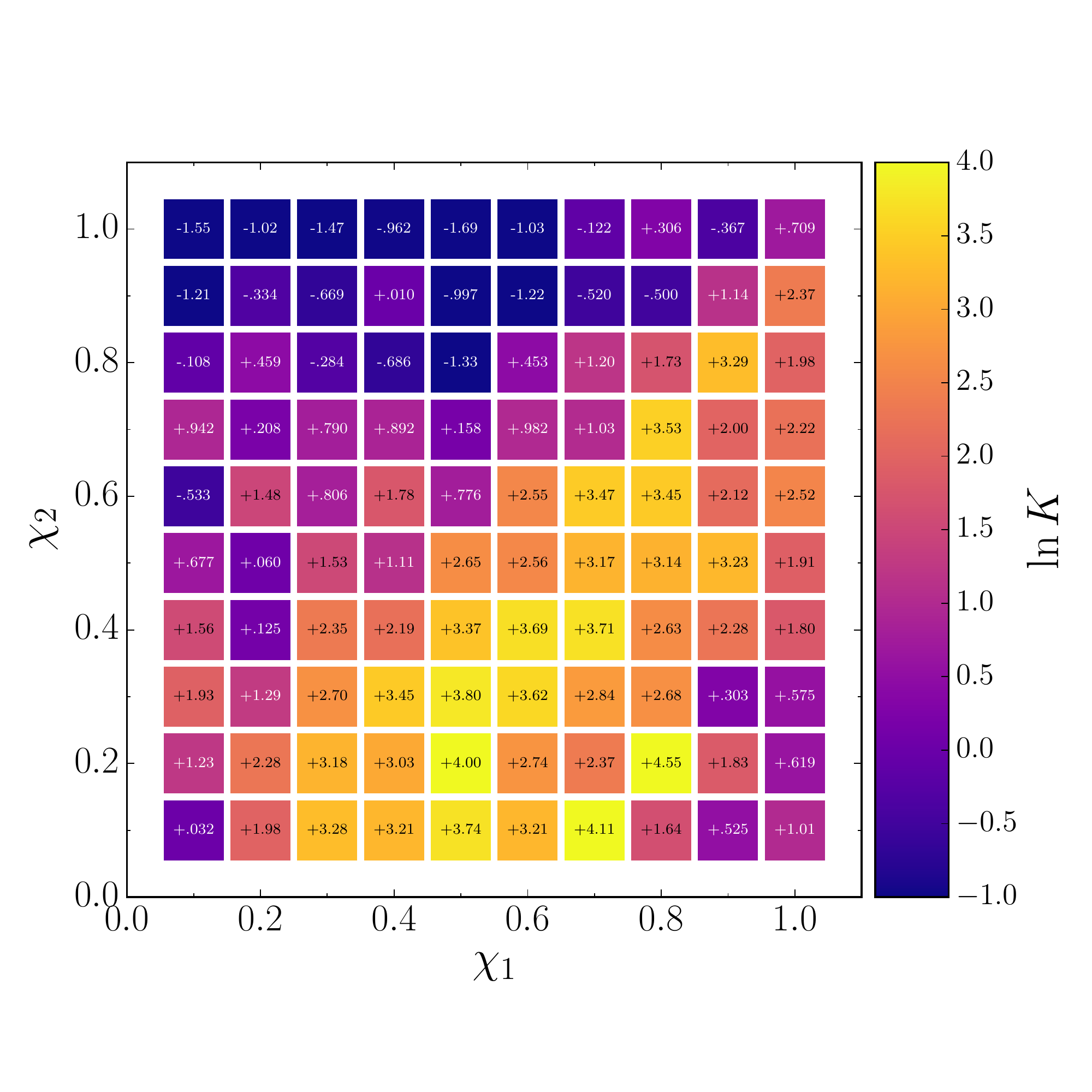}
\caption{\label{fig:Results4:TideAsymmetry}\emph{Bayes factor versus spin, with and without tides (O1)}:  For the M14
  model ($\sigma=200\unit{km/s}$), a plot of the Bayes factor versus $\chi_{1,2}$.   Colors and numbers indicate the
  Bayes factor; dark colors denote particularly unlikely configurations.   The left panel assumes no spin realignment (``no tides''); the
  right panel assumes the second-born BH's progenitor had its spin aligned with its orbit just prior to birth
  (``tides'').    Spin-orbit realignment and the high orbital velocity just prior to the second SN ensures the second
  spin is at best weakly misaligned; therefore, $\chi_2$ would need to be small for these models to be consistent with
  LIGO's observations to date.
}
\end{figure*}

Tides also introduce an asymmetry between the spin-orbit misalignment of the  first-born (generally more massive) and second-born (generally less
massive) BH \cite{2013PhRvD..87j4028G}.  As a result, when we consider general prescriptions for BH natal spins $\chi_1\ne \chi_2$, we find
that scenarios without tides produce largely symmetric constraints on $\chi_{1,2}$.   When we assume tidal
alignment, we can draw stronger constraints about the second-born spin rather than the first.  Paradoxically, large natal spin on the
\emph{first} born BH is consistent with observations.   The second born BH cannot significantly misalign its
spin through a natal kick; therefore, for comparable mass binaries like GW150914, we know that the second-born BH spin
must be small, if it is strongly aligned.   More broadly, since observations rule out large $\chi_{\rm eff}$, binary
formation scenarios with tides and  with $\chi_1>\chi_2$ fit the data substantially better
than scenarios with tides and $\chi_{2}>\chi_1$.  Because tides act to realign the second spin, only when $\chi_2\le \chi_1$ will
we have a chance at producing small $|\chi_{\rm eff} |$, as LIGO's O1 observations suggest.  
Fig. \ref{fig:Results4:TideAsymmetry} illustrates this asymmetry.  

The illustrative results described in this section follow from our strong prior assumptions:  fixed BH natal spins.  As described below,
if we instead adopt some broad distribution of BH natal spins, the substantially greater freedom to reproduce LIGO's
observations reduces  our ability to draw other distinctions, in direct proportion to the complexity of the prior hypotheses
explored.   We  describe results with more generic spin distributions below.  %

\subsection{BH natal spins, given misalignment (O1)}
\label{sec:sub:SpinDistributionIssues}

So far,  to emphasize the information content in the data, we have adopted the simplifying assumption that each
pair of BHs has the same natal spins $\chi_1,\chi_2$.    This extremely strong family of assumptions allows us to leverage all four
observations, producing large changes in Bayes factor as we change our assumptions about (all) BH natal spins.    Conversely, if the BH natal spins are nondeterministic, drawn from a  distribution
with support for any spin between $0$ and $1$, then manifestly only four observations cannot hope to constrain the BH
natal spin distribution, even were LIGO's measurements to be perfectly informative about each BH's properties.  
Astrophysically-motivated or data-driven  prior assumptions must be adopted in order to draw stronger conclusions %
about BH spins (cf. \cite{2017PhRvL.119y1103V}). %

\begin{figure*}
\includegraphics[width=0.6\textwidth]{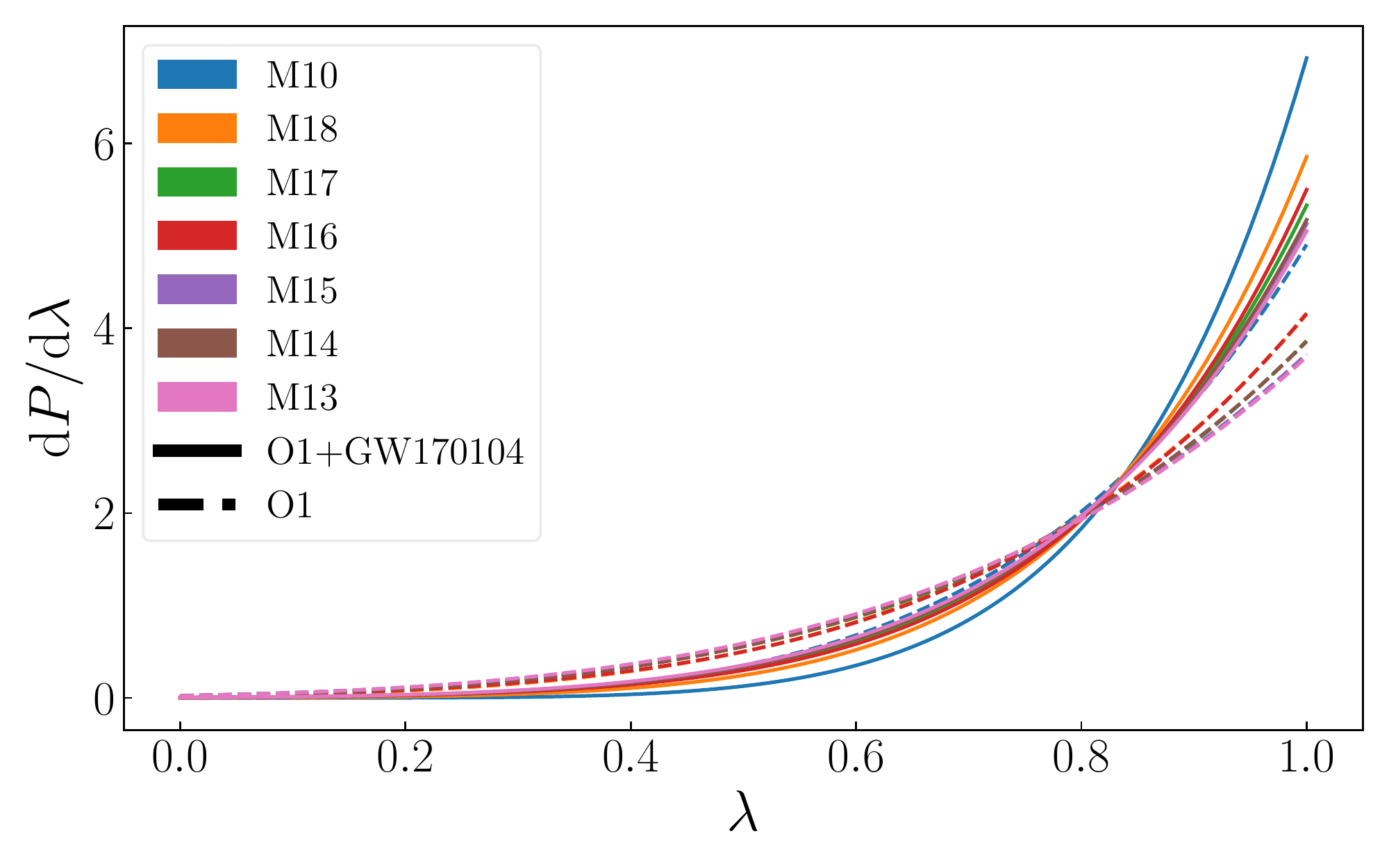}

\includegraphics[width=\columnwidth]{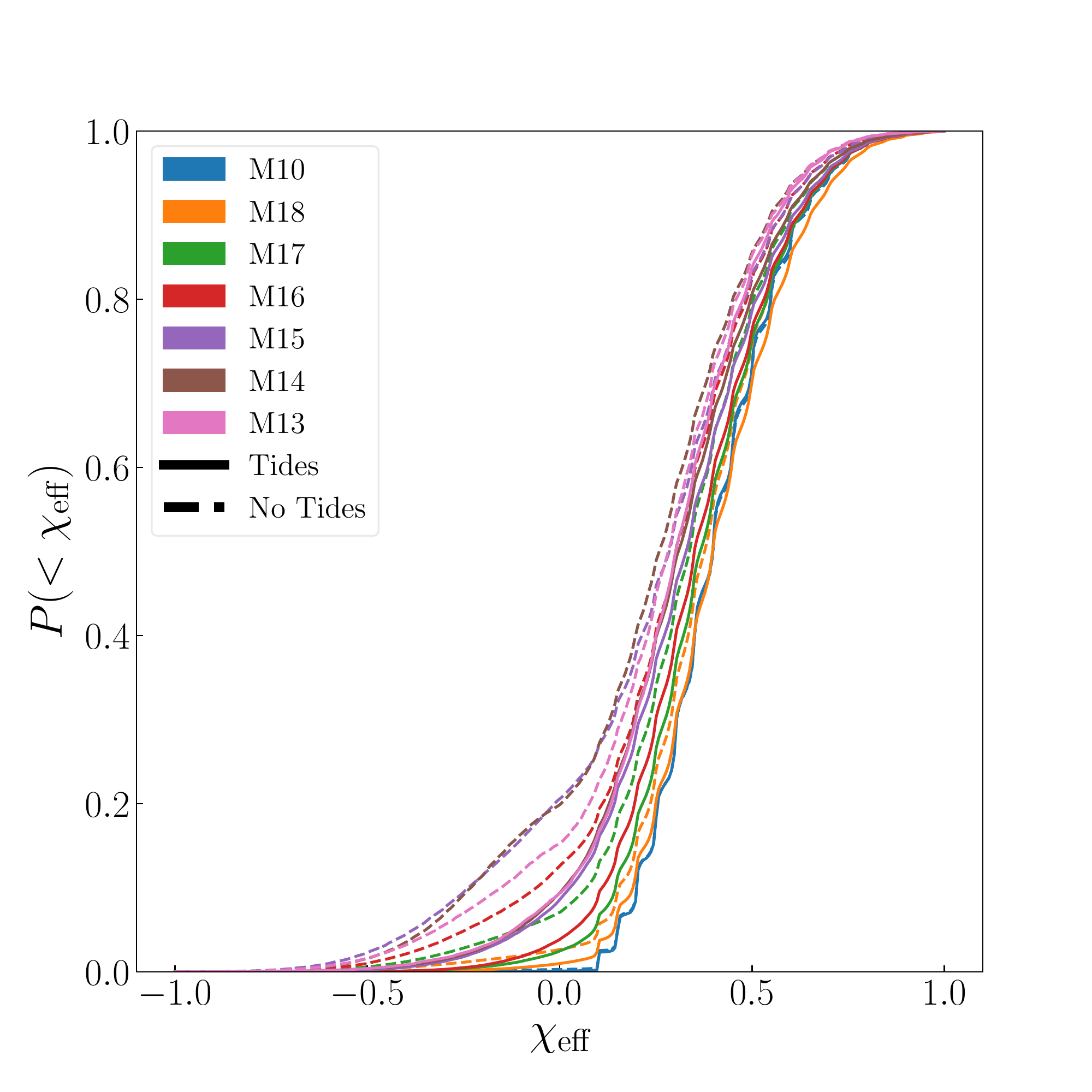}
\includegraphics[width=\columnwidth]{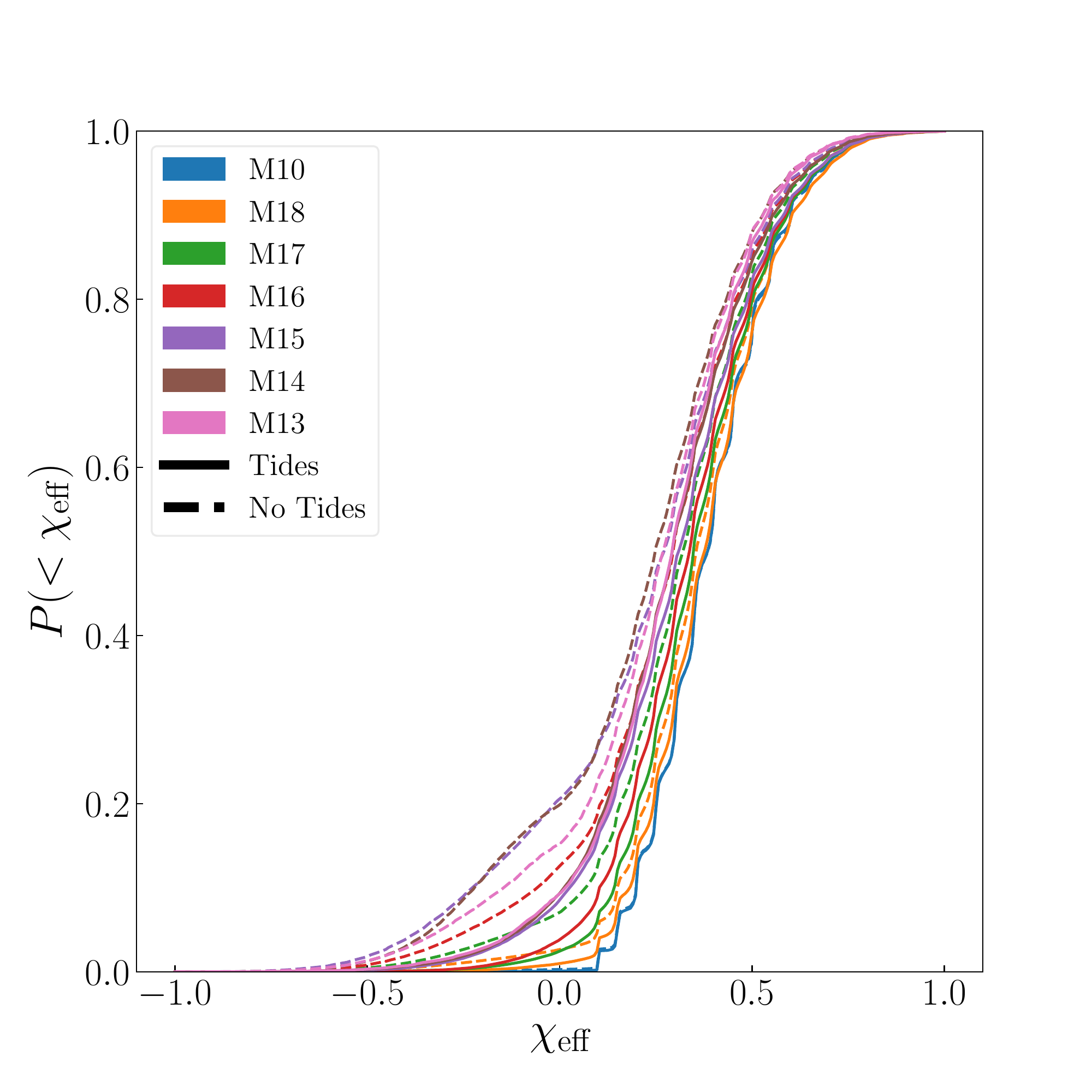}
\caption{\label{fig:PosteriorTwoBlockModel}
  \emph{High or low natal spin?}
  Top panel: Posterior distribution on $\lambda_A$, the fraction of BHs with natal spins
  $\le 0.6$ [Eq. (\ref{eq:TwoBinSpinModel})], based on O1 (dotted) or on O1 with GW170104 (solid),  compared with our binary evolution models  (colors),  assuming ``no tides.''
  Unlike
  Fig. \ref{fig:Results5:VersusChiKick}, which illustrates Bayes factors calculated assuming fixed BH natal spins, this
  calculation assumes  each BH natal spin is  drawn at random from a mass- and formation-scenario-independent
  distribution that is piecewise constant above and below $\chi=0.6$.   With only four observations,  LIGO's observations consistently
 but weakly  favor low BH natal spins. 
 Left panel: Posterior distribution for $\chi_{\rm eff}$ implied by the distribution of
 $\lambda_A$ shown in the top panel (i.e., by comparing our models to LIGO's O1 observations, under the assumptions made
 in Eq. (\ref{eq:TwoBinSpinModel})).  
  Right panel: As in the left panel, but including GW170104.  Adding this event does not appreciably or qualitatively change
  our conclusions relative to O1.
}
\end{figure*}

As a concrete example, we  consider the simple two-bin BH natal spin model described in Eq. (\ref{eq:TwoBinSpinModel}), with  probability
$\lambda_A$ that any BH has natal spin $\chi_i\le 0.6$ and probability $1-\lambda_A $ that any BH natal spin is larger
than $0.6$.   The choice of $0.6$ is motivated by our previous results in Fig. \ref{fig:Results5:VersusChiKick}, as well
as by the empirical $\chi_{\rm eff}$ distribution shown in Fig. \ref{fig:ChiEffEmpiricalCDF}. 
Using the techniques described in  Appendix \ref{ap:HierarchicalCalculation}, we can evaluate the posterior
probability for $\lambda_A$ given LIGO's O1 observations, within the context of each of our  binary evolution
models.  Fig. \ref{fig:PosteriorTwoBlockModel} shows the result: LIGO's observations weakly favor low BH natal spins.
For models like M10 and M13,  with minimal BH natal kicks and hence spin-orbit misalignment, low BH natal spin is
necessary to reconcile models with the fact that LIGO hasn't seen BH-BH binaries with large, aligned spins and thus
large $\chi_{\rm eff}$.  Conversely,  LIGO's observations will modestly less strongly
disfavor  models that frequently predict large BH natal spins (e.g., $\lambda \lesssim 0.6$).

As we increase the complexity of our prior assumptions, our ability to draw  conclusions from only four observations
rapidly decreases.    For example, we can construct the posterior distribution for a generic BH natal spin distribution
(i.e., our mixture coefficients $\lambda_\alpha$ for each spin combination can take on any value whatsoever).   The mean
spin distribution can be evaluated using closed-form expressions provided in Appendix \ref{ap:HierarchicalCalculation}.  In  this extreme case, the posterior
distribution closely resembles the prior for almost all models, except M10.

To facilitate exploration of alternative assumptions about natal spins and kicks, we have made publicly available all of the  marginalized
likelihoods evaluated in this work, as Supplemental Material \cite{supplement}. %

\subsection{Information provided by  GW170104 }
The observation of GW170104 enables us to modestly sharpen all of the  conclusions drawn above, due to the reported
limits on $\chi_{\rm eff}$: between $-0.42$ and $0.09$ \cite{2017PhRvL.118v1101A}.  %
Of course, the reported limits for all events must always be taken in context, as they are inferred using very specific assumptions---\emph{a priori} uniform spin
magnitudes, isotropically oriented.  Necessarily, inference performed in the context of any  astrophysical model
for natal BH spins and kicks will draw different conclusions about the allowed range, since the choice of prior
influences the posterior spin distribution
(see, e.g., \cite{2017PhRvD..96l4041W,2017PhRvL.119y1103V}).  %
Even taking these limits at face
value, however, this one observation can easily be explained using some combination of two effects: a significant
probability for small natal BH spins,
or some BH natal kicks.  
First and most self evidently, if all BHs have similar natal spins, then binary evolution models that assume alignment
at birth;  do not include
processes that can misalign heavy BH spins, like M10; and which adopt a common natal BH spin for all BHs  are difficult to reconcile with
LIGO's observations.
On the one hand, GW170104 would require extremely small natal spins in this scenario; on the
other, GW151226 requires nonzero spin.  
Of course, a probabilistic (mixture) model allowing for a wide range of mass-independent BH natal spins  can easily
reproduce LIGO's observations, even without permitting any alignment; see also \cite{2017arXiv170607053B},
which adopts a deterministic model that also matches these two events.  
Second, binary evolution models with significant BH natal kicks can also explain LIGO's observations.  As seen in the bottom
left panel of Fig.  \ref{fig:Results5:VersusChiKick},  large BH natal spins are harder to reconcile with LIGO's observations, if we
assume BH spin alignments are only influenced by isotropic BH natal kicks.  This conclusion follows from the modest
$\chi_{\rm eff}$ seen so far for all events.
Conversely, if we assume efficient alignment of the second-born BH, then the observed distribution of $\chi_{\rm eff}$
(and $\theta_1$, mostly for  GW151226) suggest large BH natal kicks, as illustrated by the bottom right panel of Fig. \ref{fig:Results5:VersusChiKick}.

\subsection{Information provided by the mass distribution }

\begin{figure}
\includegraphics[width=\columnwidth]{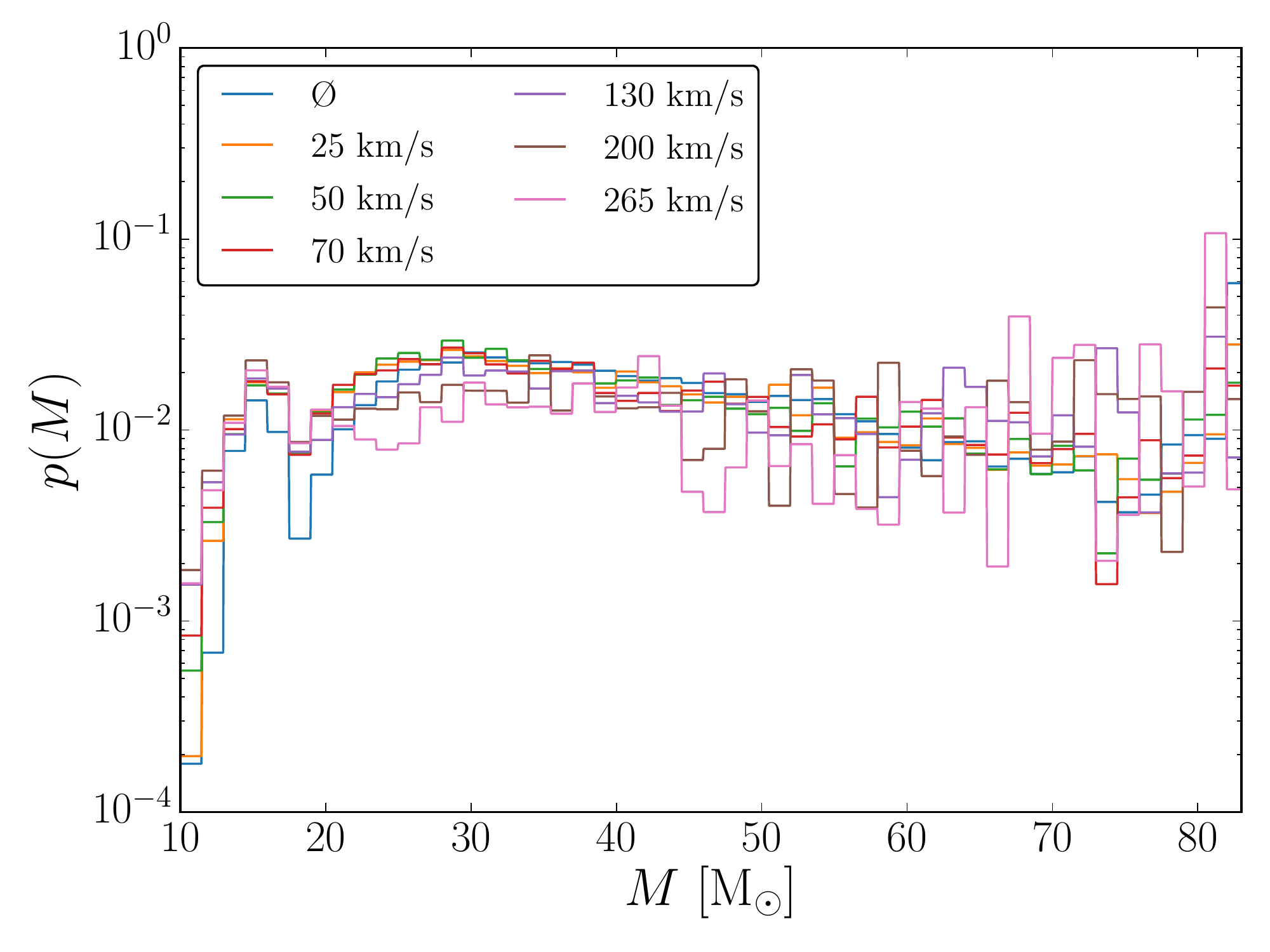}
\caption{\label{fig:MassDistributions}Detection-weighted total mass distributions of our models, labeled by their
  $\sigma_{\mathrm{kick}}$ values,  without accounting
  for LIGO measurement error. 
The overall mass distributions are very similar, particularly for low
kicks. 
}
\end{figure}

The underlying mass distributions predicted by our formation models do depend on our assumptions about BH natal kicks,
as shown concretely in   
Fig. \ref{fig:MassDistributions}.   %
These modest differences accumulate as BH natal kicks increasingly disrupt and
deplete all BH-BH binaries. 
To quantify the similarity between our
distributions, Table \ref{tab:Parameters}
reports an information-theory-based metric (the KL divergence) that attempts to
quantify the information rate or ``channel capacity'' by which the universe communicates information about the mass
distribution to us.  If $p(x), q(x)$ are two probability distributions over a parameter $x$, then in general the KL divergence has
the form %
\begin{eqnarray}
\label{eq:def:KL}
D_{KL} (p|q) = \int dx p(x)\ln [p(x)/q(x)]
\end{eqnarray}
   Except for the strongest BH natal kicks, we find our mass distributions are nearly identical.
Even with perfect mass measurement accuracy, we would
need $\mathcal{O}(1/D_{KL})$ fair draws from our distribution to confidently distinguish between them.  
As demonstrated by previous studies \cite{2015ApJ...810...58S,2017ApJ...846...82Z}, LIGO will be relatively inefficient at
discriminating between the different detected mass distributions.  LIGO is most sensitive to the heaviest  BHs, which
dominate the astrophysically observed population, but has
extremely large measurement uncertainty in this regime.  Thus, accounting for selection bias and smoothing using
estimated measurement error, the mass distributions considered here look fairly similar \cite{2015ApJ...810...58S}. 
For constraints on BH natal kicks, the  information provided by the mass distribution is far less informative than the
insights implied by constraints on $\chi_{\rm eff}$ and $\theta_{1,2}$.

As a  measure of the information LIGO can extract per event about the mass distribution from each detection, we
enumerate how many different BH-BH binaries LIGO can distinguish, which are consistent with the expected stellar-mass
BH-BH population (i.e., motivated by LIGO's reported observations to date, limiting to $m_2/m_1>0.5$, $m_1+m_2<75 M_\odot$,
$m_2>3 M_\odot$, and $m_1<40 M_\odot$).  Counting up the distinct waveforms used by gravitational wave searches in O2
\cite{2016PhRvD..93l4007C}, %
including spin, there are only \textbf{236}
templates with chirp masses above LVT151012 (i.e., $\mc>15 M_\odot$), and only $\simeq$ \textbf{1,200} with chirp masses above GW151226 (i.e.,
$\mc> 8.88 M_\odot$).
This estimate is highly optimistic, because it neglects distance and hence redshift uncertainty, which decreases our
ability to resolve the smallest masses (i.e., the uncertainty in chirp mass for GW151226), and it also uses both mass
and spin information.  
Judging from the reported mass distributions alone (e.g., the top left panel of Fig. 4 in \cite{2016PhRvX...6d1015A}), 
LIGO may efficiently isolate BHs to only a few tens of distinct mass bins, de facto limiting the resolution  of
any mass distribution which can be nonparametrically resolved with small-number statistics; see, e.g., the discussion in
\cite{2017ApJ...846...82Z}. 

\section{Predictions and projections}
\label{sec:FuturePredictions}

Using the Bayes factors  derived above for our  binary evolution models and BH natal spin assumptions (collectively
indexed by $\Lambda$), we
can make predictions about future  LIGO observations, characterized by a probability distribution  $p_{\rm future}(x) =
\sum_\Lambda p(x|\Lambda) p(\Lambda|d)$ for a candidate future binary with parameters $x$.
We can then account for  LIGO's mass-dependent sensitivity to generate the relative probability of observing binaries with
those parameters.  In the context of the infrastructure described above, we evaluate this detection-weighted posterior
probability using a mixture of  synthetic universes, with relative probabilities $p(\Lambda|d)$ and relative weight
$r_i$ of detecting an individual binary drawn from it.

Using our fiducial assumptions about BH spin realignment (``no tides''),
our posterior probabilities point to nonzero BH
natal kicks, with BH natal spins that can neither be too large nor too small (Figs. \ref{fig:Results5:VersusChiKick}
and \ref{fig:PosteriorTwoBlockModel}).  
In turn, because each of our individual formation scenarios $\Lambda$ preferentially forms binaries with $\chi_{\rm eff}>0$ \cite{DavidePaper},  with a strong preference for the largest $\chi_{\rm eff}$ allowed, we predict future LIGO observations will frequently include binaries
with the largest $\chi_{\rm eff}$ allowed by the BH natal spin distribution. %
These measurements will self-evidently allow us to constrain the natal spin distribution
(e.g., the maximum  natal BH spin).   For example, if future observations  continue to prefer  small $\chi_{\rm eff}$,
then the data would increasingly
require smaller and smaller natal BH spins, within the context of our models.  For example, this future scenario would let us
rule out  models with large kicks and large spins, as
then LIGO should nonetheless frequently detect binaries with large $\chi_{\rm eff}$.  %

As previously noted, with only four GW observations, the data does not strongly favor any spin magnitude distribution.  
Strongly modeled approaches which assume specific relationships
between the relative prior probability of different natal spins can  draw sharper constraints, as in \cite{2017Natur.548..426F}.  If we allow  the spin
distribution  to take on any form
\cite{2017MNRAS.471.2801S,2017PhRvD..96b3012T},  many  observations
would be required to draw
conclusions about the spin distribution.     
Conversely, as described previously and illustrated by Fig. \ref{fig:PosteriorTwoBlockModel}, if we adopt a weak
(piecewise-constant) model, we can draw some weak
conclusions about the BH natal spin distributions that are implied by our binary evolution calculations and LIGO's observations.
Neither the expected number of events nor their mass distribution merits extensive discussion.  
The large Poisson error implied by only four observations leads to a wide range of probable event rates, previously
shown to be consistent with all the binary evolution models presented here \cite{2016Natur.534..512B,2016A&A...594A..97B}.   Conversely, 
  due to the limited size of our model space---the discrete model set and single model parameter (BH natal kicks)
explored---these posterior distributions by no means fully encompass all of our prior uncertainty in binary evolution
and all we can learn by comparing GW observations with the data.  
While our calculations illuminate how GW measurements will inform  our understanding of BH formation, our calculations are not
comprehensive enough to provide authoritative constraints except for the most robust features.

Finally, all of our calculations and projections have been performed in the context of one family of formation scenarios---isolated binary
evolution.  Our calculations within this framework do not allow for one possible variant of this channel: homogeneous chemical evolution, where close binaries become tidally locked and rapidly rotating, leading to a distinctively different evolutionary trajectory that produces massive BH binaries while circumventing the common envelope phase \cite{2016MNRAS.458.2634M,2016A&A...588A..50M}.
Globular clusters could also produce a population of merging compact binaries \cite{2016ApJ...824L...8R},
with random spin-orbit misalignments \cite{richard-pro-forma-IsotropicVsAlignedCite}.  Several previous studies have
described or demonstrated how to identify whether either model contributes to the detected population, and by how much,
using constraints on merging BH-BH spins~\cite{2016ApJ...818L..22A,2016ApJ...832L...2R,2017CQGra..34cLT01V,2017MNRAS.471.2801S,2017PhRvL.119a1101O,2017PhRvD..96b3012T}.

\section{Conclusions}
\label{sec:Conclude}
By comparing binary evolution models with different assumptions about BH natal kicks to LIGO observations of binary BHs,
we estimate that heavy  BHs should receive a natal kick of order 50 km/s if no processes act to realign
stellar spins.  Significantly larger natal kicks, with one-dimensional velocity dispersion $\simeq 200\unit{km/s}$,  will be required if stellar spins efficiently
realign prior to the second BH's birth. 
These estimates are consistent with observations of galactic X-ray binary misalignment
\cite{1995Natur.375..464H,1997ApJ...477..876O,2001ApJ...555..489O,2010MNRAS.401.1514M} and recoil velocity
\cite{2012MNRAS.425.2799R,2015MNRAS.453.3341R,2017MNRAS.467..298R,2016ApJ...819..108B,2010MNRAS.401.1514M,2012ApJ...747..111W,2014ApJ...790..119W}.
Our estimate is driven by two simple factors. 
The natal kick dispersion $\sigma$ is bounded from above because large kicks disrupt too many binaries (reducing the
merger rate below the observed value).  
Conversely, the natal kick distribution is bounded from below because modest kicks are needed to produce a range of
spin-orbit misalignments.   A distribution of misalignments increases  our models'  compatibility  with  LIGO's observations,
if all BHs are likely to have natal spins.

Closely related work by \citet{2017arXiv170607053B} uses similar evolutionary models but with a fixed physically-motivated BH natal spin model that depends on BH mass.  They predict a distribution of $\chi_{\rm eff}$ with substantial support at large values, in increasing tension with observations reported to date.  They conclude that more efficient angular momentum transport neeeds to be adopted in evolutionary calculations to revise their BH natal spin model and to match LIGO/Virgo observations. In this work, by contrast, we explore a wide range of possible spin distributions.  We consistently find that distributions which favor low BH natal spins can more easily reproduce current observations.

Given limited statistics, we have for simplicity (and modulo M10) assumed all binary BHs receive   natal kicks and spins drawn from the
same formation-channel-independent distributions. 
This strong assumption about BH natal spins allows us to draw sharp inferences about BH natal spins and kicks by combining complementary information provided by
GW151226 (i.e., nonzero spins required, with a suggestion of misalignment) and the remaining LIGO observations (i.e.,
strong limits on $\chi_{\rm eff}$).   
Future observations will allow us to directly test more complicated models not explored here, where the natal spin and kick distribution depends on the
binary BH mass as in \citet{2017arXiv170607053B} 
Necessarily, if BH natal spins are  small for massive BHs and large for small BHs, as proposed in
\citet{2017arXiv170607053B}, then  measurements of low-mass BH
binaries like GW151226 will provide our primary channel into constraining  BH natal spins and kicks.
At present, however, inferences about BH natal spins and spin-orbit misalignment are strongly model or equivalently
prior driven, with sharp conclusions only possible with strong assumptions.  We strongly recommend results about future
BH-BH observations be reported
or interpreted using multiple and astrophysically motivated priors, to minimize confusion about their astrophysical
implications (e.g., drawn from the distribution of $\chi_{\rm eff}$).

For simplicity, we have also only adjusted one assumption (BH natal kicks) in our  fiducial model for how compact
binaries form.  A few other pieces of unknown and currently parametrized physics, notably the physics of common envelope evolution, should play a
substantial role in how compact binaries form and, potentially, on BH spin misalignment.  Other assumptions have much smaller impact on the event rate and particularly on BH spin misalignment.  
Adding additional sources of uncertainty will generally diminish the sharpness of our conclusions.  For example, the net
event rate depends on the assumed initial mass function as well as the star formation history and metallicity
distribution throughout the universe; once all systematic uncertainties in these inputs are inclusded,
the relationship between our models and the expected number of events is likely to include significant systematic as
well as statistical uncertainty.   Thus, after marginalizing over all sources of uncertainty, the event rate may not be
as strongly discriminating between formation scenarios. 
By employing several independent observables (rate, masses, spins
and misalignments), each providing weak constraints about BH natal kicks, we protect our conclusions against systematic
errors in the event rate.   Further investigations are needed to more fully assess sources of systematic error and enable more precise constraints.

Due to the limited size of our model space---the discrete model set and single model parameter (BH natal kicks) explored---these posterior distributions by no means fully encompass all of our prior uncertainty in binary evolution and all we can learn by comparing GW observations with the data.  As in previous early work \cite{2005ApJ...633.1076O,2008ApJ...672..479O,2008ApJ...675..566O,2010CQGra..27k4007M},
a fair comparison must broadly explore many more
elements of uncertain physics in binary evolution, like mass transfer and stellar winds.  
 Nonetheless, this  nontrivial example of astrophysical inference shows how we can learn about
astrophysical models via simultaneously comparing  GW measurements of several parameters  of several detected binary BHs
to predictions of any model(s).  While we have applied our statistical techniques to  isolated binary evolution, these tools can be
applied to generic formation scenarios, including homogeneous chemical evolution; dynamical formation in globular
clusters or AGN disks; or even primordial binary BHs. %

Forthcoming high-precision astrometry and radial velocity from GAIA will enable higher-precision constraints on existing X-ray binary proper motions
and distances \cite{2016A&A...595A...1G,2016A&A...595A...2G}, as well as increasing the sample size of available BH binaries. These forthcoming improved constraints on BH binary
 velocities  will provide a complementary avenue to constrain BH natal kicks using binaries in our own
 galaxy.

\begin{acknowledgements}
We thank Christopher Berry, Simon Stevenson, and Will Farr for helpful comments on the draft.
D.W. is supported by the Rochester Institute of Technology (RIT) through the Frontiers in Gravitational Wave Astrophysics (FGWA) Signature Interdisciplinary Research Areas (SIRA) initiative and College of Science.
R.O. is supported by NSF Grants No. AST-1412449, PHY-1505629, and PHY-1607520.
D.G. is supported by NASA through Einstein Postdoctoral Fellowship Grant No. PF6-170152 awarded by the Chandra X-ray Center, which is operated by the Smithsonian Astrophysical Observatory for NASA under Contract NAS8-03060.
E.B. is supported by NSF Grants No. PHY-1607130 and AST-1716715, and by FCT contract
IF/00797/2014/CP1214/CT0012 under the IF2014 Programme.
M.K. is supported by the Alfred P. Sloan Foundation Grant No. FG-2015-65299 and NSF Grant No. PHY-1607031.
R.O. and E.B. acknowledge the hospitality of the Aspen Center for Physics, supported by NSF PHY-1607611, where this work was completed.
K.B. acknowledges support from the Polish National Science Center (NCN) grant: Sonata Bis 2 (DEC-2012/07/E/ST9/01360).
D.E.H. was partially supported by NSF CAREER grant PHY-1151836 and NSF Grant No. PHY-1708081. He was also supported by the Kavli Institute for Cosmological Physics at the University of Chicago through NSF Grant No. PHY-1125897 and an endowment from the Kavli Foundation.
Computations were performed on the Caltech computer cluster ``Wheeler,'' supported by the Sherman Fairchild Foundation and Caltech. Partial support is acknowledged by NSF CAREER Award PHY-1151197.
The authors thank to the LIGO Scientific Collaboration for access to the data and gratefully acknowledge the support of the United States National Science Foundation (NSF) for the construction and operation of the LIGO Laboratory and Advanced LIGO as well as the Science and Technology Facilities Council (STFC) of the United Kingdom, and the Max-Planck-Society (MPS) for support of the construction of Advanced LIGO. Additional support for Advanced LIGO was provided by the Australian Research Council.

\end{acknowledgements}

\appendix

\section{\uppercase{Approximating parameter distributions from finite samples}}
\label{ap:gmm-approximation}

Our population synthesis techniques allow us to generate an arbitrarily high number of distinct binary evolutions from each formation scenario,
henceforth indexed by  $\Lambda$.  Instead of generating individual binary evolution histories, we weigh each one by an
occurrence rate, allowing it to represent multiple binaries. 
For our calculations, however, we instead require the relative probability of different binaries, not just samples from
the distribution.  
We estimate
this distribution from the large but finite sample of binaries available in each synthetic universe.   We do not simply
use an occurrence rate-weighted histogram of all the samples. 
  Histograms work reliably for any single parameter (e.g., $p(m_1 | \Lambda)$), where many samples are available per
  potential histogram bin, but for high-dimensional
joint distributions (e.g., $p(m_1, m_2, \theta_1, \theta_2, \chi_{\mathrm{eff}} | \Lambda)$), many histogram bins will
be empty simply due to the curse of dimensionality. 

In all our calculations, we instead  approximate the density as a mixture of Gaussians, 
labeled $k = 1,2,\ldots,K$, with means and covariances ($\boldsymbol\mu_k$,
$\boldsymbol\Sigma_k$) to be estimated, along with weighting coefficients $w_k$, which must sum to unity. The density
can therefore be written as 
\begin{equation}
  p(\boldsymbol{x}) \approx
  \sum_{k=1}^K
    w_k \, \mathcal{N}(\boldsymbol{x} | \boldsymbol\mu_k, \boldsymbol\Sigma_k),
  \label{eq:gmm-density}
\end{equation}
where $\mathcal{N}(\cdot)$ represents the (multivariate) Gaussian distribution.
We select the number of Gaussians $K$ by using the Bayesian information criterion. %

To estimate the means and covariances of our mixture of Gaussians, we used
the expectation maximization algorithm \cite{EMAlgorithm}; see, e.g.,  \cite{murphy2012machine} for a pedagogical
introduction.  
Specifically, we used a small modification to an implementation in  \texttt{scikit-learn} \cite{scikit-learn}, to allow
for weighted samples in the update equation (e.g., adding weights to Eq. (11.27) in \cite{murphy2012machine}).  

Ideally we would simply approximate each formation scenario $\Lambda$'s intrinsic predictions $p(\mathbf{x}|\Lambda)$
with a mixture of Gaussians, using the merger rate for each sample binary as its weighting factor. However, all
astrophysical indications suggest that more massive progenitors form more rarely, implying this
procedure would result in a distribution that is strongly skewed in favor of the much more intrinsically frequent low
mass systems;  our fitting algorithm might end
up effectively neglecting the samples with small weights. This would risk losing information about the most
observationally pertinent samples, which due to LIGO's mass-dependent sensitivity are concentrated at the highest
observationally accessible masses.   
Alternatively, for every choice of detection network, we can approximate each formation scenario's predictions \emph{for
  that network}.  If $TV(\mathbf{x})$ is the average sensitive 4-volume for the network, according to  this procedure we
approximate $V(\mathbf{x})p(\mathbf{x})$ by a Gaussian mixture, then divide by $V(\mathbf{x})$ to estimate $p(\mathbf{x})$.  
To minimize duplication of effort involved in regenerating our approximation for each detector network, we instead adopt
a \emph{fiducial} (approximate) network sensitivity model $V_{\rm ref}(\mathbf{x})$ for the purposes of density
estimation.    %
  We adopt
the simplest  (albeit ad-hoc) network sensitivity model: the functional form for $V(\mathbf{x})$ that arises by
using a single detector network and ignoring cosmology (i.e., $E{V} \propto \mathcal{M}_c^{15/6}$)
\cite{2010ApJ...716..615O}.   The overall, nominally network- and run-dependent normalization constant in this ad-hoc 
model $V_{\rm ref}$ scales out of all final results.

\section{\uppercase{Hierarchical comparisons of observations with data}}
\label{ap:HierarchicalCalculation}

As described in Sec. \ref{sec:sub:CompareMarginalLikelihood}, the population of binary mergers accessible to our
light cone can be described as an inhomogeneous Poisson process, characterized by a probability density
$e^{-\mu}\prod_k {\cal R} p(x_k)$ where $x_k = x_1\ldots x_N$ are the distinct binaries in our observationally
accessible parameter volume ${\cal V}$.  In this expression, the expected number of events and parameter distribution
are related by $\mu = \int dx \sqrt{g} {\cal R} p(x)$; the multidimensional integral $\int dx \sqrt{g}$ is shorthand for a suitable
integration over a manifold with metric; and the probability density $p(x)$ is expressed relative to the fiducial (metric)
volume element, but normalized on a larger volume than ${\cal V}$.   %
Accounting for data selection \cite{2004AIPC..735..195L}, the likelihood of all of our observations is therefore given
by Eq. (\ref{eq:Likelihood}). 

To insure we fully capture the effects of precessing spins, we work not with the full  likelihood---a difficult function
to approximate in 8 dimensions---but instead with a
fiducial  posterior distribution $p_{\rm post} = Z^{-1} p(d_k|x)p_{\rm ref}(x_k)$, as would be provided by a Bayesian calculation using a 
 reference prior $p_{\rm ref}(x_k)$.   Rewriting the integrals $\int dx_k p(d_k |x) p(x_k|\Lambda)$ appearing in Eq. (\ref{eq:Likelihood})  using the reference prior
we find integrals appearing in this expression can be calculated by Monte Carlo, using some sampling distribution
$p_{s,k}(x_k)$ for each event (see, e.g., \cite{2010ApJ...725.2166H}):
\begin{align}
 \int dx_k & p(d_k|x_k) p(x_k|\Lambda) \nonumber \\
 &= \frac{1}{N_{k}} 
   \sum_s 
   \frac{[p(d_k|x_k) p_{\rm ref}(x_{k,s})]p(x_{k,s}|\Lambda)
    }{
    p_{s,k}(x_{k,s}) p_{\rm ref}(x_{k,s})},
\end{align}
where $s=1,\ldots N_{k}$ indexes the Monte Carlo samples used.   One way to evaluate this integral is to adopt a
sampling distribution $p_{s,k}$ equal to the posterior distribution evaluated using the reference prior, and thus proportional to $p(d_k|x_k)p_{\rm ref}(x_k|\Lambda)$.
If for this event $k$ we have samples $x_{k,s}$ from the posterior   distribution---for example, as provided by a Bayesian Markov chain Monte Carlo code---the integrals appearing in Eq. (\ref{eq:Likelihood}) can be estimated by
\begin{eqnarray}
\int dx_k  p(d_k|x_k) p(x_k|\Lambda) \simeq  \frac{Z}{ N_{k}}\sum_s \frac{p(x_{k,s}|\Lambda)}{p_{\rm ref}(x_{k,s})},
\end{eqnarray}
We use this expression to evaluate the necessary marginal likelihoods, for any proposed observed population  $p(x|\Lambda)$.
In the expression above, we need only consider \emph{some} of the degrees of freedom in the problem. 
Notably, the probability distributions for extrinsic parameters like the source orientation, sky location, and distance
will always be in common between our models and our reference prior.  So will any Jacobians associated with changes of coordinate.  Moreover, these assumptions are independent of one
another and of the intrinsic parameter distributions.  Therefore, the ratio of probability densities $p(x|\Lambda)/p_{\rm ref}(x)$ usually has product form, canceling term by term.   We therefore truncate the ratio to only account for
\emph{some} of the degrees of freedom.

To verify and better understand our results, we can also \emph{approximate} the likelihood function, using
suitable summary statistics.  %
 As an example, \citet{2016PhRvD..94f4035A} reproduce parameter estimates of
GW150914 using a Gaussian approximation to the likelihood and the assumption of perfect spin-orbit alignment.    Using
this approximation, and a similar approximation for GW151226, we can alternatively approximate each integral appearing
in the likelihood by using the (weighted) binary evolution samples $x_{k,A}$  and their weights $w_A$:
\begin{eqnarray}
\int dx_k  p(d_k|x_k) p(x_k|\Lambda) \simeq  \frac{\sum_A  w_A  \hat{p}(d_k|x_{k,A})}{\sum_A w_A}
\end{eqnarray}
where $\hat{p}$ refers to our approximate likelihood for the $k$th event.   
Even though these likelihood approximations neglect degrees of freedom associated with spin precession, we can reproduce
the observed mass and $\chi_{\rm eff}$ distributions reported in \citet{2016PhRvD..94f4035A}.   We used this approximate
likelihood approach
 to validate and test our procedure.  We also use this approach to incorporate information about GW170104,
 which was not available at the same level of detail as the other events.  
As an example, we describe how to evaluate this integral in the
case where $p(x_k|\lambda)$ is a mixture model
$p(x|\lambda) = \sum_\alpha \lambda_\alpha p_\alpha(x)$, for $\lambda$ an array of parameters.  In this case, all the
integrals can be carried out via
\begin{align}
\label{eq:Likelihood:Mixture}\
\prod_k & \int dx_k  p(d_k|x_k)p(x_k|\lambda)    \\
= &\prod_k [\sum_\alpha \lambda_\alpha \int dx_k p(d_k |x_k)p_\alpha(x_k)] %
 = \prod_k \sum_\alpha \lambda_\alpha c_{\alpha,k} \nonumber
\end{align}
where $c_{\alpha,k}$ are integrals we can compute once and for all  for each event, using for example the posterior
samples from some fiducial analysis.  
As a result, the observation-dependent factor in likelihood for a mixture model always reduces to a homogeneous
$N$th-degree polynomial in the mixture parameters $\lambda_\alpha$.   %
Bayes theorem can be applied to $\lambda$ to infer the distribution over mixture parameters.  Depending on the mixture
used, this calculation could incorporate a physically-motivated prior on $\lambda$.  %

We use a mixture model approach to hierarchically constrain  the spin magnitude  distribution implied by our data.  In
our approach, we first consider models where both spin magnitudes are fully constrained.  In the notation of the mixture
model discussion above, we adopt some specific prior
$p_{\alpha}(\chi_1,\chi_2|\sigma) = \delta(\chi_1 -x_\alpha)\delta(\chi_2-y_\alpha)$ where $x_\alpha,y_\alpha$ are the spin
$\lambda_\alpha$.   A mixture model allowing generic $\lambda$ and thus including all such components  allows both
component spins to take arbitrary  (discrete) values.   [We could similarly extend our mixture model to
include kicks.] %
The posterior distribution over all possible  spin distributions  $p(\lambda|d) = p(d|\lambda)p(\lambda)/p(d)$  follows from Bayes'
theorem and the concrete likelihood given in Eq. (\ref{eq:Likelihood:Mixture}).   In practice, however, we don't
generally compute or report the full posterior distribution, as it contains far more information than we need (e.g., the
extent of the ensemble of possible spin distributions that fit the data).    Instead, we compute the \emph{expected spin
  distribution}
\begin{eqnarray}
p_{\rm post}(x) = \sum_\alpha \E{\lambda_\alpha}p_\alpha(x)
\end{eqnarray}
and the \emph{variance in each $\lambda_\alpha$}.  
For the modest number  of mixture components of interest here ($\simeq 100$ possible choices of both spin magnitudes)
and the modest degree of the polynomial ($\simeq 4-5$),
all necessary averages can be computed by direct symbolic quadrature of a polynomial in $\lambda_\alpha$.   The integral
can be expressed as a sum of terms of homogeneous degree in $\lambda$, and integrals of each of these terms can be carried out via the following general formula:
\begin{widetext}
\begin{eqnarray}
n! \int_{\sum_i x_i \le 1} dx_1 \ldots dx_n x_{i_1}^{\alpha_1} \ldots x_{i_Z}^{\alpha_Z} = \frac{n!}{(n-Z)!} \prod_{k=1}^Z B(\alpha_k+1, n+1-k +\sum_{q>k} \alpha_q)
\end{eqnarray}
\end{widetext}
where the integral is over the region $x_i\ge 0$ and $\sum_i x_i\le 1$.
We can also find the maximum likelihood estimate of $\lambda_\alpha$, for example by using the expectation-maximization
algorithm \cite{EMAlgorithm}. 
In this work, however, we have many more basis models $\alpha=1,2,\ldots $ used in our (spin) mixture  than
observations.  Normally, we would reduce the effective dimension, for example by adopting prior assumptions in how the
mixture coefficients can change as a function of spins $\chi_{1},\chi_2$.   To minimize additional formal overhead, we instead
simply treat the spins hierarchically in blocks [Eq. (\ref{eq:TwoBinSpinModel})], considering lower-dimensional models where (for example) $\lambda_A$
denotes the \emph{a priori} probability  for $\chi_{i} \le 0.6$  and $1-\lambda_A$ denotes the \emph{a priori} probability for
$\chi_i>0.6$, so for example the prior probability for $(\chi_1,\chi_2)=0.1$ is $\lambda_A^2/36$. 
In this four-block and one-parameter model, we can compute the average value of $\lambda_A$ in terms of the net weights
associated with each block:
$C_{AA,k}=\sum_{\chi_1,\chi_2 \in A} c_{\alpha,k}$, $C_{A\bar{A}k}$, $C_{\bar{A}Ak}$ and $C_{\bar{A}\bar{A}k}$.   For
example, if for each of three synthetic observations, $C_{AA}=1$ and all other weights are negligible, then we would
conclude \emph{a posterori} that  $\E{\lambda_A}=0.875$ and $\sigma_{\lambda_A}=0.11$.  
This approach was adopted in Fig. \ref{fig:PosteriorTwoBlockModel}, in contrast to the preceding figures  which
adopted fixed natal spins for all BHs.

\section{\uppercase{Approximate posterior distribution for GW170104}}
\label{ap:mockup-GW170104}

For most events examined in this study, we made use of  posterior samples provided and performed by the LIGO Scientific
Collaboration, generated by comparing  each event to  the IMRPv2 approximation \cite{2014PhRvL.113o1101H}. Because we cannot employ the same
level of detail %
for GW170104, we instead resort to an approximate posterior distribution, derived from the reported
GW170104 results \cite{2017PhRvL.118v1101A} and our understanding of gravitational wave parameter estimation, as
approximated using a Fisher matrix \cite{1994PhRvD..49.2658C}.

For GW170104 we construct an approximate (truncated) Gaussian posterior distribution in only three
correlated binary parameters: $\mc,\eta,\chi_{\rm eff}$.  The shape  of this Gaussian (i.e., its inverse covariance
matrix) was constructed via a Fisher matrix approximation, derived using the median detector-frame parameters reported for GW170104
(i.e., $m_{1}\simeq 37.1 M_\odot$, $m_2 \simeq 22.6 M_\odot$, and---breaking degeneracy with an ad-hoc choice---$\chi_{1,z} \chi_{2,z} \simeq \chi_{\rm eff} \simeq -0.12$);  the reported network SNR of GW170104 (i.e., $\rho\simeq
13.0$); and a suitable single-detector noise power spectrum.  Our effective Fisher matrix estimate for the inverse
covariance matrix $\Gamma$ \cite{2013PhRvD..87b4004C}  adopted the noise power spectrum at
GW150914, using a minimum frequency $f_{\rm min}=30\unit{Hz}$;  employed the (nonprecessing) SEOBNRv4 approximation \cite{2017PhRvD..95d4028B},
evaluated on a grid in $\mc,\eta, \chi_{1,z},\chi_{2,z}$; and fit as a quadratic function of $\mc,\eta,\chi_{\rm eff}$.  
We adopt a nonprecessing model and lower-dimensional Fisher matrix approximation because the posterior of this event,
like GW150914, is consistent with nonprecessing spins and is very well approximated, in these parameters, by a
nonprecessing model; see, e.g., \cite{2016PhRvD..94f4035A}.  
This simple approximation captures important correlations between $\mc,\eta$ and $\chi_{\rm eff}$, and
the diagonal terms of $\Gamma^{-1}\rho^2$ roughly reproduce the width of the posterior distribution reported for
GW170104.  To obtain better agreement with the reported one-dimensional credible intervals,  we scaled
the terms  $\Gamma_{\mc,x}$ for $x=\mc,\eta,\chi_{\rm eff}$ by a common scale factor $0.29$ and the term
$\Gamma_{\chi_{\rm eff},\chi_{\rm eff}}$ by $0.9$.
For similar reasons, we likewise  hand-tuned the center of the Gaussian distribution to the (unphysical) parameter location
to $\mathcal{M}_{\mathrm{c}} = 22.9$, $\eta = 0.32$, $\chi_{\mathrm{eff}} = 0.013$.
Using this ansatz, we generate  GW170104-like posterior samples in $\mc,\eta,\chi_{\rm eff}$ from this Gaussian
distribution, truncating any unphysical samples (i.e., with $\eta>1/4$).   For our tuned posterior,  the median and $90\%$ credible
regions on the synthetic posteriors approximate the values and ranges reported.   According to our highly
simplified and purely synthetic approach, the resulting $90\%$ credible regions are $M_{\mathrm{tot}} = 51.2^{+7.6}_{-6.8} M_\odot$, $q = 0.62^{+0.25}_{-0.24}$,
$\chi_{\mathrm{eff}} = -0.12^{+0.28}_{-0.27}$.

\bibliography{paper}

\end{document}